\documentclass[floatfix, twocolumn,superscriptaddress,10pt]{revtex4-2}
\usepackage[T1]{fontenc}
\usepackage{times}
\usepackage[utf8]{inputenc}
\usepackage[resetlabels,labeled]{multibib}
\usepackage{amsmath}
\usepackage{amsthm}
\usepackage{amsfonts}
\usepackage{amssymb}
\usepackage{bm}
\usepackage{bbm}
\usepackage{bbold}
\usepackage{graphicx,epsfig,color,tcolorbox}
\usepackage{float}
\usepackage{physics}
\usepackage{mathtools}
\usepackage{mathrsfs}
\usepackage{latexsym}
\usepackage[colorlinks=true,linkcolor=teal,citecolor=purple,breaklinks=true]{hyperref}
\usepackage[ruled,vlined]{algorithm2e}
\usepackage[mathscr]{eucal}
\usepackage{eucal}

\newcommand{\ot}{\otimes}
\newtheorem{lemma}{Lemma}

\definecolor{plbred}{rgb}{0.8,0,0}

\newcommand{\marti}{\color{black}}
\newcommand{\chg}[1]{\color{black}#1 \color{black}}

\begin{document}

\title{Operational Definition of the Temperature of a Quantum State}
\author{Patryk Lipka-Bartosik}
 \affiliation{Department of Applied Physics, University of Geneva, 1211 Geneva, Switzerland}
\author{Mart\'i Perarnau-Llobet}
 \affiliation{Department of Applied Physics, University of Geneva, 1211 Geneva, Switzerland}
 \author{Nicolas Brunner}
  \affiliation{Department of Applied Physics, University of Geneva, 1211 Geneva, Switzerland}

\date{\today}

\begin{abstract}
Temperature is usually defined for physical systems at thermal equilibrium. Nevertheless one may wonder if it would be possible to attribute a meaningful notion of temperature to an arbitrary quantum state, beyond simply the thermal (Gibbs) state. In this work, we propose such a notion of temperature considering an operational task, inspired by the Zeroth Law of thermodynamics. Specifically, we define two effective temperatures for quantifying the ability of a quantum system to cool down or heat up a thermal environment. In this way we can associate an operationally meaningful notion of temperature to any quantum density matrix. We provide general expressions for these effective temperatures, for both single- and many-copy systems, establishing connections to concepts previously discussed in the literature. Finally, we consider a more sophisticated scenario where the heat exchange between the system and the thermal environment is assisted by a quantum reference frame. This leads to an effect of ``coherent quantum catalysis'', where the use of a coherent catalyst allows for exploiting quantum energetic coherences in the system, now leading to much colder or hotter effective temperatures. \chg{We demonstrate our findings using a two-level atom coupled to a single mode of the electromagnetic field.}
\end{abstract}

\maketitle

\section{Introduction}
Temperature is a well-defined  property of macroscopic systems in thermal equilibrium~\cite{callenthermodynamics}. When considering  equilibrium systems of finite size, subtleties on the notion of temperature can arise due to the breakdown  of the equivalence of statistical  ensembles  \cite{Touchette2009,Hilbert2014,Mller2015,brandao2015equivalence,Campisi2015a,Hnggi2016,campisi2021lectures} and the non-negligible effect of interactions between constituents~\cite{Hartmann2004,Ferraro_2012,Kliesch2014,Hern_ndez_Santana_2015,Hern_ndez_Santana_2021}. 
Moving on to non-equilibrium (quantum) systems, assigning an effective temperature can be useful in certain physical contexts~\cite{casas2003temperature,popov2007ontology,Puglisi2017,Hsiang_2021}. Given some quantum state $\rho$ evolving under Hamiltonian $H$, the most common approach is to assign an effective temperature $T^*$ given by the temperature of an equilibrium (Gibbs) state with the same average energy, that is
\begin{align}
\Tr[H \rho] = {\rm Tr}[H \gamma(T^*, H)]
\label{eq:T*}
\end{align}
where $\gamma(T, H)\equiv e^{-\beta H}/Z$, $Z = \Tr e^{-\beta H}$ is the partition function and $\beta=1/k_B T$ ($k_B\equiv1$). The relevance of the identification \eqref{eq:T*} naturally arises in the dynamics of isolated quantum many-body systems: seminal results suggest that $\rho$ will become in practice indistinguishable \footnote{More precisely, they become indistinguishable for physically relevant local observables and under mild conditions for $\rho$ and $H$, see e.g. counterexamples in integrable systems \cite{kinoshita2006quantum,Brenes2020} and many-body localisation \cite{anderson1958absence,Abanin_2019,nandkishore2015many}} from $\gamma(T^*,H)$ after a transient thermalisation time~\cite{Deutsch1991,Srednicki1994,Rigol2008,Polkovnikov2011,Eisert2015,Gogolin2016,Mitchison2022}. Beyond isolated systems, different notions of effective temperatures have been discussed for characterizing non-equilibrium states \cite{Schnack1998,Alipour2021}, and also in the context of quantum thermal machines \cite{Brunner2012,skrzypczyk2015passivity,Silva2016,Mitchison2019,Rossnagel2014,Abah_2014,correa2014quantum}.

Here we propose an alternative and operational approach for defining temperature for quantum systems. We consider a system with Hamiltonian $H$ in a quantum state $\rho$, and place it in contact with another (reference) system initially at thermal equilibrium. We then assign two temperatures for the system, $T_c$ and $T_h$, which correspond to the lowest and the highest temperatures at which the reference system can be cooled down or heated up. That is, $T_c$ and $T_h$ characterize the potential of $\rho$ to heat up or cool down a reference state in  thermal equilibrium (see Fig.~\ref{fig:1}). This approach establishes a natural direction of heat flow between  $\rho$ and a thermal environment at temperature  $T$:  Heat will always flow  towards the environment when $T\leq T_c$ and, likewise, the environment will always \chg{release} heat when $T\geq T_h$. Instead, for $T_c\leq T \leq T_h$, the direction of heat flow depends on the particular process. As expected, $T_c\leq T^* \leq T_h$, with equality when the system itself is in thermal equilibrium with temperature $T^*$.

We apply this approach to three different situations. First, we find {\marti{explicit expressions for the effective temperatures of quantum systems, which are related to the concept of virtual temperatures~\cite{Brunner2012,skrzypczyk2015passivity, Mitchison2019,Silva2016,Janzing2000,Quan2007,Koukoulekidis2021}}}. Second, we show that for macroscopic quantum systems, the two effective temperatures are closely related to $T^*$. Third, we extend the framework by introducing a reference frame, or a quantum catalyst~\cite{Jonathan_1999,Brand_o_2015,Mueller2018,Ng_2015,Wilming2017,shiraishi2021quantum,wilming2021entropy,Korzekwa2022}. This gives access to strictly colder and hotter effective temperatures by exploiting energy coherences present in the system.

\section{Results}
\subsection{Operational definition of temperature}
A system $S$ will be described by a tuple $(H_S, \rho_S)$, where $H_S$ stands for the system's Hamiltonian and $\rho_S$ its density matrix. We say that a system is in thermal equilibrium at temperature $T$ if its state can be written as $\rho_S = \gamma(T, H_S)$. 

\begin{figure}
    \centering
    \includegraphics[width=\linewidth]{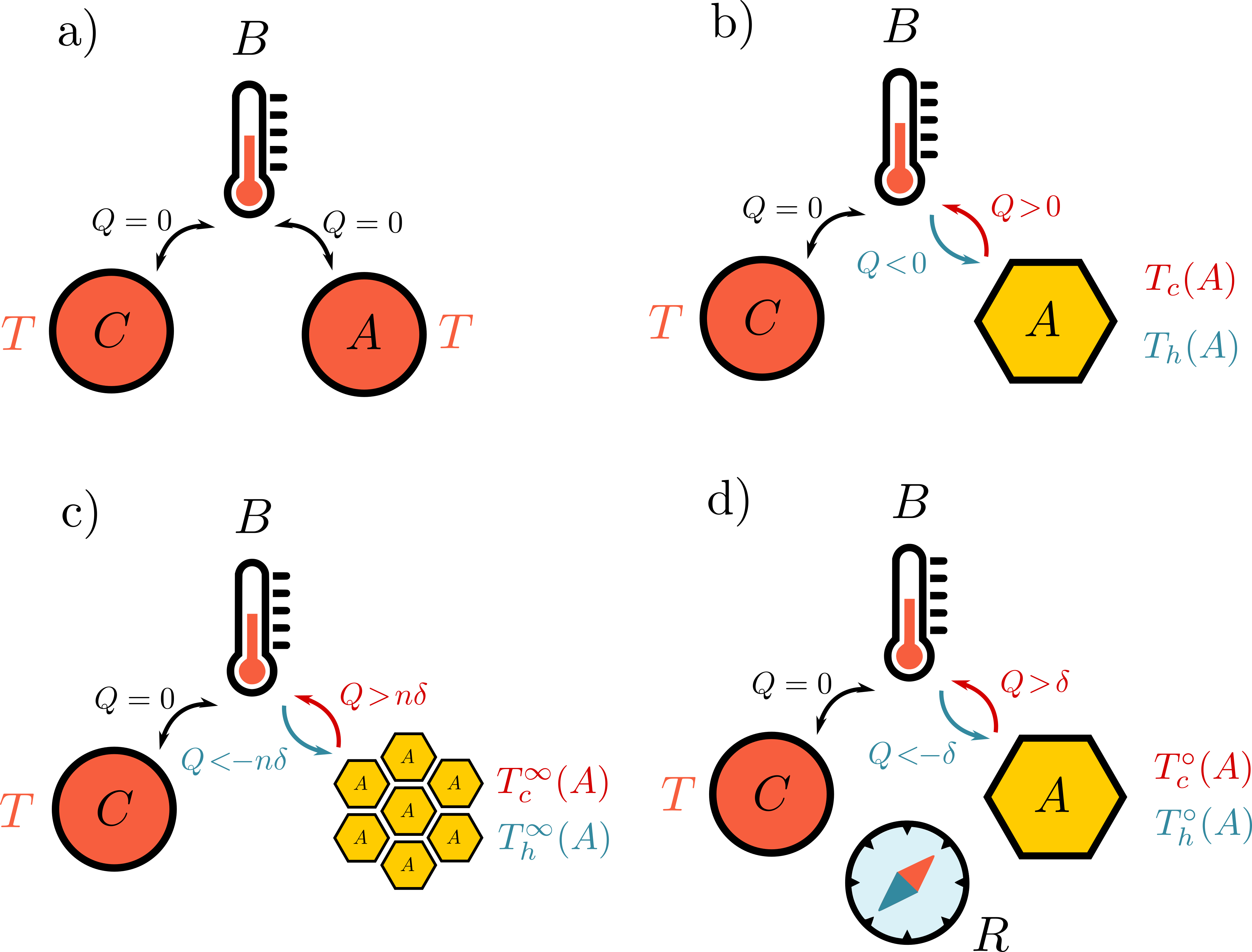}
    \caption{Setups used to define effective temperatures for non-equilibrium quantum systems. Fig. $(a)$ shows the equilibrium setting addressed by the Zeroth Law. Fig. $(b)$ shows the setup used to assign effective cold $T_c(A)$ and hot $T_h(A)$ temperatures with system $A$. Fig. $(c)$ describes the case when effective temperatures are assigned to multiple copies of $A$. In Fig. $(d)$ we extend the setting by adding a reference frame (catalyst), i.e. a system that aids the process without providing heat itself. For details see the main text.}
    \label{fig:1}
\end{figure}

The Zeroth Law of thermodynamics states that when system~$A$ is in a thermal equilibrium with another system~$B$ that is in a thermal equilibrium with~$C$, then~$A$ must be in a thermal equilibrium with~$C$ (see Fig. \ref{fig:1}a) \cite{callen1998thermodynamics,lieb1999physics}. Importantly, the Zeroth Law associates temperature with systems in thermal equilibrium. Here we will use a similar approach to define \emph{effective temperatures} for non-equilibrium systems. We will consider three systems: $A, B$ and $C$, where $C$ is a macroscopic heat bath at temperature $T$, system $B = (H_B, \gamma(T, H_B))$ is a thermometer probe, and $A = (H_A, \rho_A)$ is the quantum system whose effective temperatures we want to quantify. For that, we consider the following steps: $(i)$   we couple $B$ to $C$ until $B$ reaches thermal equilibrium at temperature $T$, $(ii)$  we decouple them, and $(iii)$  we couple $B$ to $A$ to infer the effective temperature of $A$ by measuring $B$.

Let us now describe in detail step $(iii)$. First, we demand that energy is preserved within the joint system $AB$, i.e. the evolution is described by a unitary process $U$ satisfying $[U, H_A + H_B] = 0$. This ensures that only the energy of $A$ is used to heat up or cool down the thermometer. Second, we assume perfect control over the joint system, meaning we allow for arbitrary processes $U$. The dynamics is then characterized by the full set of energy-preserving unitaries \cite{Janzing2000,DAlessandro2007}. Notice that we make no assumptions about the strength of interaction (weak or strong), its complexity (local or collective), nor duration (short or long) with respect to the natural time scales. Third, we assume the initial state of $AB$ factorizes, i.e. $\rho_{AB} = \rho_{A} \ot \gamma_B(T, H_B)$, ensuring that $B$ has a well-defined temperature and its energy changes may be interpreted as heat. 

The joint state of the system and the thermometer after the interaction is given by $\sigma_{AB} = U\left[ \rho_A \ot \gamma_B(T, H_B) \right]U^{\dagger}$. The heat transferred to the thermometer is therefore
\begin{align}
     Q(T, H_B, U)\! :=\! \Tr\left[H_B (\sigma_B - \gamma_B(T, H_B))\right],
\end{align}
with $\sigma_B:=\Tr_A\sigma_{AB}$. It is well-known that heat can only flow in one direction when $A$ is  a Gibbs state at some temperature~$T(A)$, i.e.~$\rho_A = \gamma_A(T(A), H_A)$. More specifically, 

\begin{align}
\label{eq:equilibriumcase}
 & Q(T,H_B,U)
   \geq 0 \hspace{5mm}  \quad \text{for all} \quad   T\leq T(A), \nonumber\\
   & Q(T,H_B,U) \leq 0 \hspace{5mm}  \quad \text{for all} \quad T \geq T(A),  
\end{align}
which holds for all possible $H_B$ and $U$~\cite{Loyd19898,jennings2010entanglement}. 
In other words, the direction of heat is well-defined between Gibbs states: Heat can only flow from hot to cold, regardless of the particular thermometer $B$ or process $U$. 
Instead, for non-thermal states the direction of heat flow is not always unique, as illustrated by its reversal in the presence of correlations~\cite{Loyd19898,jennings2010entanglement,micadei2019reversing,Henao_2018}. That is, the sign of $Q(T,H_B,U)$ in general depends on $H_B$ and the process~$U$. 

The crucial insight of our work is that there are temperatures for which  heat  has a well-defined direction even for non-equilibrium states (i.e., it is independent of~$H_B$ and~$U$ and depends only on~$A$). Given some $A=(H_A,\rho_A)$,  we will show that there exist temperatures $T_c(A)$ and $T_h(A)$ that satisfy 
\begin{align}
\label{eq:nonequilibriumcase}
 & Q(T,H_B,U)
   \geq 0 \hspace{5mm}  \quad \text{for all} \quad   T \leq T_c(A), \nonumber\\
   & Q(T,H_B,U) \leq 0 \hspace{5mm}  \quad \text{for all} \quad   T \geq T_h(A),  
\end{align} 
for all possible $H_B$ and $U$,
in  analogy with the equilibrium case~\eqref{eq:equilibriumcase}. 
As a consequence, $A$ is effectively hotter/colder than any equilibrium state below/above $T_c(A)$ and $T_h(A)$, respectively. Conversely, $A$~has the \emph{potential} to cool down equilibrium states with temperature  $T\geq T_c(A) $ and heat up states with $T\leq T_h(A)$. Hence,  $T_c(A)$ and $T_h(A)$ bound the ability of $A$ to heat up or cool down a thermal environment. 

In order to find the effective cold $T_c(A)$ and hot $T_h(A)$ temperatures, we look for the minimal (maximal) temperature $T$ for which $Q(T,H_B, U)$ is negative (positive) for \emph{some} thermometer $B$ and \emph{some} energy-preserving interaction $U$, i.e.
\begin{equation}
\begin{aligned}
\label{eq:beta_c_def}
T_c(A) :=\quad \min_{H_B, U} \quad & T  \\
\textrm{s.t.} \quad & Q(T, H_B, U) < 0, \\
\quad & [U, H_A+H_B] = 0.
\end{aligned}
\end{equation}
Similarly, we define $T_h(A)$ by replacing $\max$ with $\min$ above and reversing the inequality.  

\subsection{Effective temperatures of single-copy quantum systems}
The protocol discussed above is effectively described using a channel acting on the system $A$:
\begin{align}
   \label{eq:4}
    \mathcal{E}_{T}(\cdot) =\, &\Tr_B[U(\cdot \ot \gamma_B(T, H_B))U^{\dagger}].
\end{align}
This class of channels is known as \emph{thermal operations} \cite{Janzing2000,Horodecki2013,Brand_o_2015,Lostaglio2019,Korzekwa2016,Chubb_2018}. We use this correspondence to prove our first main result. More specifically, for \emph{any} quantum system $A = (H_A, \rho_A)$ with $H_A = \sum_{i} \epsilon_i \dyad{\epsilon_i} $ and $p_i := \bra{\epsilon_i}\rho_A\ket{\epsilon_i}$, we show that the effective temperatures are given by
\begin{align}
    \label{eq:teff_vts}
   T_c(A) = \min_{i \neq j}\,\, T_{ij}, \qquad
   T_h(A) = \max_{i \neq j}\,\, T_{ij},
\end{align}
where \chg{$T_{ij} := (\epsilon_j - \epsilon_i)/\log(p_i/p_j)$} {\marti{are the virtual temperatures of the state} \cite{Janzing2000,Quan2007,Brunner2012,Silva2016,skrzypczyk2015passivity,Mitchison2019,Koukoulekidis2021}}. \chg{As expected, for Gibbs states $T_c(A)=T_h(A)$, whereas for  non-equilibrium states $T_c(A) \neq T_h(A)$.} The proofs proceed similarly for both effective temperatures, so we only discuss $T_c(A)$. First, we show that for any system $A$ there is an interaction and a thermometer $B$ that can be cooled down when $T \geq \min_{i \neq j}\,\, T_{ij}$. {\marti The cooling protocol is simple: Consider the two-dimensional subspace $\{\ket{\epsilon_k},\ket{\epsilon_l}\}$ corresponding to the lowest virtual temperature, i.e.  $T_c(A)=T_{kl}$. We then introduce a two-level thermometer $B$ with levels $\{\ket{g},\ket{e}\}$ and energy gap $\epsilon_k-\epsilon_l$, thus in resonance with the selected subspace. By coupling such subspaces via an energy-conserving operation $U = e^{-i H_{\text{int}}t}$ acting for time $t = \pi/2$ with $H_{\text{int}} = \dyad{\epsilon_l}{\epsilon_k}_A \ot \dyad{g}{e}_B + \text{h.c.}$, the energy of $A$ will necesseraly increase and hence $B$ will be cooled down, as desired. Second,  we use the resource-theoretic approach to show that when $T \leq \min_{i \neq j}\,\, T_{ij}$, there exists no  protocol that can cool down any thermometer $B$.   
This proves that $T_c(A)$ is equal to $\min_{i \neq j}\,\, T_{ij}$. For details see Appendix~\ref{app:1}.
}

The effective temperatures can be both positive and negative. This is a consequence of their operational character: A system with a negative effective temperature has its energy population inverted (note that the effective temperature \eqref{eq:T*} also becomes negative in this case). Therefore, a quantum system with $T_h(A) < 0$ can heat up  equilibrium systems at any real temperature $T$. Similarly, when $T_c(A) < 0$, the system cannot cool down  any equilibrium system, even when its real temperature is arbitrarily large. 

A quantum system $A$ is out of equilibrium with respect to  temperature $T$ when it contains at least one virtual temperature different from $T$. Moreover, since virtual temperatures depend only on the occupations in the energy basis, i.e. on $p_i = \bra{\epsilon_i}\rho_A\ket{\epsilon_i}$, superpositions of energy levels have the same ability to generate heat as corresponding probabilistic mixtures.  {\marti This means that energy coherences  are irrelevant from the perspective of cooling or heating the thermometer --which is intimately connected to the time-translation symmetry of the allowed operations~\cite{lostaglio2015description,lostaglio2015quantum,Korzekwa2016,Marvian2019} }
We now discuss two ways to overcome this restriction. First, we consider processing multiple copies of the system collectively~\cite{Alicki2013,salvia2020energy,salvia2022extracting}  and show that the degeneracy of energy levels allows to exploit coherence locked in the quantum system. Second, we extend the framework by introducing a reference frame \chg{(or a catalyst)}, i.e. a system that allows to locally lift some of the restrictions imposed by the presence of conserved quantities. 

\begin{figure}
    \centering
    \includegraphics[width=\linewidth]{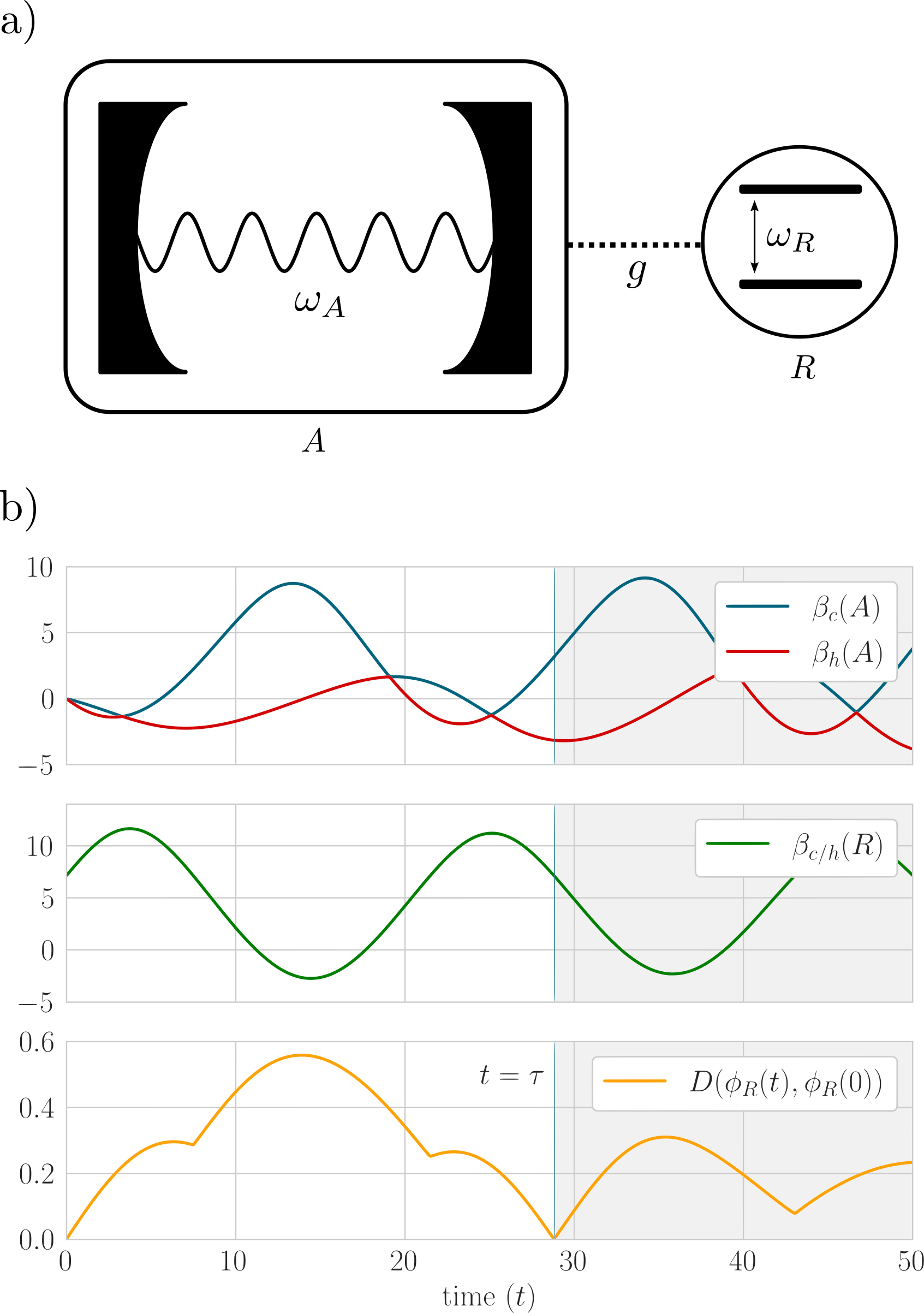}
    \caption{The illustrative example described in the main text. Panel $(a)$ shows the sketch of the scenario: a two-level system $R$ interacting resonantly with a single-mode optical cavity $A$. Panel $(b)$ describes the time evolution of the effective temperatures of both systems. The top figure shows effective temperatures of $A$ as a function of time $t$. The middle panel shows the evolution of the effective temperatures of $R$. The bottom panel shows the distance between the states of the atom at some time $t$ and $t = 0$ as quantified by the trace distance $D(\rho, \sigma) := \Tr[\sqrt{(\rho-\sigma)^{\dagger}(\rho-\sigma)}]/2$. Notice that at time $t = \tau$  the system $R$ returns to its initial state, therefore it acts as a catalyst.  The parameters chosen are $\omega_R = \omega_A = 1$, $g = 0.1$ and $\tau = 28.5$.}
    \label{fig:4}
\end{figure}

\subsection{Effective temperatures of macroscopic quantum systems} 

Suppose that system $A$ consists of multiple identical copies, i.e. $A \equiv A^n := (H_A^{\ot n}, \rho_A^{\ot n})$ (see Fig. \ref{fig:1}c). {\marti As we will see, it is then convenient to  extend the} definition of effective temperatures \eqref{eq:beta_c_def} by introducing a parameter $\delta > 0$ capturing the minimal amount of heat transferred to (or measured by) the thermometer, i.e.
\begin{equation}
\begin{aligned}
\label{eq:beta_c_delta_def}
T_c(A, \delta) :=\quad \min_{H_B, U} \quad & T  \\
\textrm{s.t.} \quad & Q(T, H_B, U) \leq -\delta, \\
\quad & [U, H_A+H_B] = 0.
\end{aligned}
\end{equation}
The definition for $T_h(A,\delta)$ is obtained by replacing $\max$ with $\min$, reversing the inequality, and changing the sign of $\delta$. For consistency, we can verify that $\lim_{\delta \rightarrow 0_+} T_c(A, \delta) \equiv T_c(A)$.   For macroscopic systems comprised of $n$ particles, we require that $\delta$ is proportional to $n${\marti, so that the transferred heat is also macroscopic.} Therefore, we define the \emph{asymptotic effective temperatures} as
\begin{align}
    \label{eq:asymp_eff_temp}
    T^{\infty}_{c/h}(A,\delta) \equiv \beta^{\infty}_{c/h}(A,\delta)^{-1} := \lim_{n \rightarrow \infty} T_{c/h}(A^n, n\delta).
\end{align}
In Appendix B we show that $\beta^{\infty}_{c/h}(A,\delta)$ can be expressed as
\begin{align}
    \label{eq:beta_c_asymp}
    \beta^{\infty}_c(A,\delta) &= \frac{1}{\delta}\left[S(\gamma_A(E+\delta)) - S(\rho_A)\right], \\
    \label{eq:beta_h_asymp}
     \beta^{\infty}_h(A,\delta) &= \frac{1}{\delta}\left[S(\rho_A) - S(\gamma_A(E-\delta)) \right],
\end{align}
where $\gamma_A(x)$ stands for a Gibbs state with Hamiltonian $H_A$ and average energy $x$. Notice the change of notation for the Gibbs state introduced to simplify the formulas that follow. Furthermore, we introduced a parameter $E:=\tr(\rho_A H_A)$ and $S(\rho) := -\Tr\rho\log\rho$ is the von Neuman entropy. {\marti It can be noted that, while $T_c(A)$ and $T_h(A)$ correspond to the minimal and maximal virtual temperatures of $A$, the temperatures $T_c(A,\delta)$ and $T_h(A, \delta)$ depend on the whole spectrum of $\rho_A$ and hence all its virtual temperatures.}

To develop some intuition about the effective temperatures we can look at the thermodynamic limit. Here it corresponds to the regime with $n \rightarrow \infty$ and $\delta \rightarrow 0$ with $A$ prepared in a Gibbs state with average energy $E$, i.e. $\rho_A = \gamma_A(E)$. In this regime $\beta^{\infty}_{c/h}(A,\delta)$ both converge to the usual definition of \chg{(inverse)} temperature, i.e.
\begin{align}
    \lim_{\delta \rightarrow 0}\,\, \beta_c^{\infty}(A,\delta)  = \lim_{\delta \rightarrow 0} \,\, \beta_h^{\infty}(A,\delta)   = \frac{\partial S(\gamma_A(E))}{\partial E}.
\end{align}
For any quantum state $\rho_A$ we can further expand up to $\mathcal{O}(\delta^2)$,
\begin{align}
    \label{eq:expansion_eff_temp}
   \beta_{c/h}^{\infty}(A,\delta) \approx \pm \frac{\Delta S(\rho_A)}{\delta} + \beta^*(E) - \frac{\delta}{2 \Delta^{\!2} E(\gamma_A)},
\end{align}
where $\Delta S(\rho_A) := S(\gamma_A(E)) - S(\rho_A)$ and $\Delta^2 E(\gamma_A) := \tr[H_A^2 \gamma_A(E)] - \tr[H_A \gamma_A(E)]^2$. 
Expression \eqref{eq:expansion_eff_temp} naturally connects $ \beta_{c/h}^{\infty}(A,\delta)$ with the effective temperature $\beta^*$ defined in Eq.~\eqref{eq:T*}.
For generic states $\beta_{c/h}(A,\delta)$ differ, i.e. quantum systems can be both hot and cold, even in the asymptotic limit. Note that $\beta_{c/h}(A,\delta)$ diverge in the limit $\delta \rightarrow 0$, indicating that an asymptotically large source of non-equilibrium can cool down or heat up reference systems at any temperature  by a sublinear amount in $n$. To illustrate how the range of effective temperatures changes by considering multiple copies of~$A$, in Appendix \ref{app:3} we discuss a simple toy example.  


{\marti The effective temperatures \eqref{eq:beta_c_asymp} directly depend on~$S(\rho_A)$, which means that quantum (energy) coherences are relevant in this collective scenario, in contrast to the single-copy case.} \chg{ One way to understand this is that by considering more copies of the system, one increases the size of degenerate energy eigenspaces. This enables more flexibility in transferring population, or equivalently, generating and accepting heat using energy-conserving interactions.} {\marti   In what follows we discuss an alternative approach to achieve the same  goal. 
}

\subsection{Catalysis and the role of quantum coherence}

{\marti We now explore the possibility of exploiting}  energy coherences via the concept of catalysis~\cite{Aberg2014,Vaccaro2018,Jonathan_1999,Mueller2018,Brand_o_2015,Lipka-Bartosik2021,shiraishi2021quantum,Gallego2016,Wilming2017,boes2019bypassing,Boes2019,Ng_2015,Henao2022,Henao_2021,Marvian2019,lostaglio2019coherence,PhysRevLett.127.080502}. In this case an auxiliary system (the catalyst) provides a phase reference for the main system. Crucially, after the interaction, the catalyst must be returned to its initial state. This ensures that it provides no energy and can be later reused. This mechanism allows taking advantage of energy coherences at the level of a single copy of the system. {\marti Let us consider a simple example  to illustrate the strength of this approach. }

\chg{
Consider the evolution of a two-level atom coupled to a single mode of electromagnetic field in an optical cavity (see Fig. \ref{fig:4}). For that, let $A$ denote the cavity with bosonic creation and annihilation operators $a^{\dagger}$ and $a$. Furthermore, let $R$ be the two-level system with raising and lowering operators $\sigma_+ = \dyad{e}{g}$ and $\sigma_- = \dyad{g}{e}$. The interaction is modeled using the Jaynes-Cummings Hamiltonian \cite{Jaynes1963}, which in the rotating wave approximation reads
\begin{align}
    H_{AR} = \omega_A a^{\dagger} a + \omega_R  \dyad{e} + H_{\text{int}},
\end{align}
where $H_{\text{int}} := g \left(\sigma_+ a + \sigma_- a^{\dagger} \right)$, $\omega_{A}$ is the angular frequency of the mode and $\omega_{R}$ is the atomic transition frequency. To keep this example relatively simple we truncate the number of Fock states of $A$ to three levels \footnote{Taking more Fock states ($N$) does not change the qualitative aspects of our analysis. The only difference is that the number of different virtual temperatures will be generally proportional to $N$, leading to a more convoluted evolution picture.}. Moreover, we assume that the atom is driven on resonance, i.e. $\omega_A = \omega_R$. This ensures that the unitary evolution $U(t) = e^{-i H_{AR} t}$ generated by $H_{AR}$ satisfies $[U(t), H_{AR}] = 0$ for all times $t$. Our goal is to quantify the effective temperatures of the cavity, $T_{c/h}(A)$. 

Assume that the field $A$ and atom $R$ start in states $\ket{\psi}_A = \left(\ket{0} + \ket{1} + \ket{2}\right)/3$, $\phi_R = X(\tau)$, where $\tau$ is a free parameter and $X(\tau)$ is the state of the atom obtained by solving the operator equation
\begin{align}
    \label{eq:xt}
    X(\tau) = \Tr_{A}[U(\tau) \left(\dyad{\psi}_A \ot X(\tau) \right) U(\tau)^{\dagger}].
\end{align} 
In other words, the state of the atom is chosen so that at time $t = \tau$  returns to its initial state. We now compare two different cases: non-interacting ($g = 0$) and interacting ($g > 0$). 

When $g = 0$ both systems evolve independently. The effective temperatures $T_{c/h}(A)$ can be computed from Eq. (\ref{eq:teff_vts}) and, in our particular case, they read $    T_c(A) = T_h(A) = 0.$ 
As expected, in the absence of interaction, the presence of the atom does not influence the effective temperatures of the electromagnetic field.

When~$g > 0$, the coupling between the field~$A$ and the atom~$R$ changes their energy occupations with time, and therefore also changes their effective temperatures. After time~$t = \tau$ the atom returns to its initial state,~$\phi_R(\tau) = \phi_R(0)$, however the field may end up in a different state (see Fig \ref{fig:4}b). As a consequence, its spectrum of virtual temperatures can change{\marti, as shown in Fig \ref{fig:4}b.} 
This is a generic effect, i.e. for any value of~$\tau$ we can find the corresponding state of the atom which returns to its initial state by solving Eq. (\ref{eq:xt}). Interestingly, after time~$\tau$ the photonic mode~$A$ has lost quantum coherence, indicating a tradeoff between coherence of~$A$ and its ability to generate a flow of heat, as captured by the effective temperatures. This indicates a genuinely
quantum mechanism that exploits energy coherence to generate the desired flow of heat.   
}

We now ask how general this behavior is, i.e. what are the effective temperatures when using general catalysts? \chg{For that, we shall now consider arbitrary (energy-conserving) interactions and arbitrary states of the catalyst}. 
{\marti In Appendix~\ref{app:4}, we show that,
via a catalytic system, one can reach the same effective temperatures as in the macroscopic case of Eq.~(\ref{eq:asymp_eff_temp}). This naturally connects the two approaches considered. }
Importantly, in this case, system $A$ is microscopic, i.e. $\delta$ quantifies the total transferred heat  (rather than heat per particle). Moreover, no catalyst $R$ that remains unchanged can lead to lower cold and higher hot effective temperatures. 

\textbf{\emph{Discussion}}. We proposed an operational definition of temperature for non-equilibrium quantum systems. We defined two effective temperatures that quantify the ability of a quantum state to generate a flow of heat when coupled to a thermal environment (thermometer). We showed that these effective temperatures are given by the maximal and minimal virtual temperatures of the system and connected them with the effective temperature $T^*$ in the asymptotic limit. We then extended this setting by allowing for the use of coherent reference frames, and found that energy coherences can influence effective temperatures.

\textbf{\emph{Acknowledgements.}} We acknowledge the Swiss National Science Foundation for financial support through the Ambizione grant PZ00P2-186067 and the NCCR QSIT and NCCR SwissMAP.

\bibliographystyle{apsrev4-2}
\bibliography{references}

\begin{thebibliography}{81}%
\makeatletter
\providecommand \@ifxundefined [1]{%
 \@ifx{#1\undefined}
}%
\providecommand \@ifnum [1]{%
 \ifnum #1\expandafter \@firstoftwo
 \else \expandafter \@secondoftwo
 \fi
}%
\providecommand \@ifx [1]{%
 \ifx #1\expandafter \@firstoftwo
 \else \expandafter \@secondoftwo
 \fi
}%
\providecommand \natexlab [1]{#1}%
\providecommand \enquote  [1]{``#1''}%
\providecommand \bibnamefont  [1]{#1}%
\providecommand \bibfnamefont [1]{#1}%
\providecommand \citenamefont [1]{#1}%
\providecommand \href@noop [0]{\@secondoftwo}%
\providecommand \href [0]{\begingroup \@sanitize@url \@href}%
\providecommand \@href[1]{\@@startlink{#1}\@@href}%
\providecommand \@@href[1]{\endgroup#1\@@endlink}%
\providecommand \@sanitize@url [0]{\catcode `\\12\catcode `\$12\catcode
  `\&12\catcode `\#12\catcode `\^12\catcode `\_12\catcode `\%12\relax}%
\providecommand \@@startlink[1]{}%
\providecommand \@@endlink[0]{}%
\providecommand \url  [0]{\begingroup\@sanitize@url \@url }%
\providecommand \@url [1]{\endgroup\@href {#1}{\urlprefix }}%
\providecommand \urlprefix  [0]{URL }%
\providecommand \Eprint [0]{\href }%
\providecommand \doibase [0]{https://doi.org/}%
\providecommand \selectlanguage [0]{\@gobble}%
\providecommand \bibinfo  [0]{\@secondoftwo}%
\providecommand \bibfield  [0]{\@secondoftwo}%
\providecommand \translation [1]{[#1]}%
\providecommand \BibitemOpen [0]{}%
\providecommand \bibitemStop [0]{}%
\providecommand \bibitemNoStop [0]{.\EOS\space}%
\providecommand \EOS [0]{\spacefactor3000\relax}%
\providecommand \BibitemShut  [1]{\csname bibitem#1\endcsname}%
\let\auto@bib@innerbib\@empty
\bibitem [{\citenamefont {Callen}()}]{callenthermodynamics}%
  \BibitemOpen
  \bibfield  {author} {\bibinfo {author} {\bibfnamefont {H.}~\bibnamefont
  {Callen}},\ }\href@noop {} {\bibinfo {title} {Thermodynamics and an
  introduction to thermostatics, 1985}}\BibitemShut {NoStop}%
\bibitem [{\citenamefont {Touchette}(2009)}]{Touchette2009}%
  \BibitemOpen
  \bibfield  {author} {\bibinfo {author} {\bibfnamefont {H.}~\bibnamefont
  {Touchette}},\ }\href {https://doi.org/10.1016/j.physrep.2009.05.002}
  {\bibfield  {journal} {\bibinfo  {journal} {Physics Reports}\ }\textbf
  {\bibinfo {volume} {478}},\ \bibinfo {pages} {1} (\bibinfo {year}
  {2009})}\BibitemShut {NoStop}%
\bibitem [{\citenamefont {Hilbert}\ \emph {et~al.}(2014)\citenamefont
  {Hilbert}, \citenamefont {H\"anggi},\ and\ \citenamefont
  {Dunkel}}]{Hilbert2014}%
  \BibitemOpen
  \bibfield  {author} {\bibinfo {author} {\bibfnamefont {S.}~\bibnamefont
  {Hilbert}}, \bibinfo {author} {\bibfnamefont {P.}~\bibnamefont {H\"anggi}},\
  and\ \bibinfo {author} {\bibfnamefont {J.}~\bibnamefont {Dunkel}},\ }\href
  {https://doi.org/10.1103/PhysRevE.90.062116} {\bibfield  {journal} {\bibinfo
  {journal} {Phys. Rev. E}\ }\textbf {\bibinfo {volume} {90}},\ \bibinfo
  {pages} {062116} (\bibinfo {year} {2014})}\BibitemShut {NoStop}%
\bibitem [{\citenamefont {M\"{u}ller}\ \emph {et~al.}(2015)\citenamefont
  {M\"{u}ller}, \citenamefont {Adlam}, \citenamefont {Masanes},\ and\
  \citenamefont {Wiebe}}]{Mller2015}%
  \BibitemOpen
  \bibfield  {author} {\bibinfo {author} {\bibfnamefont {M.~P.}\ \bibnamefont
  {M\"{u}ller}}, \bibinfo {author} {\bibfnamefont {E.}~\bibnamefont {Adlam}},
  \bibinfo {author} {\bibfnamefont {L.}~\bibnamefont {Masanes}},\ and\ \bibinfo
  {author} {\bibfnamefont {N.}~\bibnamefont {Wiebe}},\ }\href
  {https://doi.org/10.1007/s00220-015-2473-y} {\bibfield  {journal} {\bibinfo
  {journal} {Communications in Mathematical Physics}\ }\textbf {\bibinfo
  {volume} {340}},\ \bibinfo {pages} {499} (\bibinfo {year}
  {2015})}\BibitemShut {NoStop}%
\bibitem [{\citenamefont {Brandao}\ and\ \citenamefont
  {Cramer}(2015)}]{brandao2015equivalence}%
  \BibitemOpen
  \bibfield  {author} {\bibinfo {author} {\bibfnamefont {F.~G.}\ \bibnamefont
  {Brandao}}\ and\ \bibinfo {author} {\bibfnamefont {M.}~\bibnamefont
  {Cramer}},\ }\href@noop {} {\bibfield  {journal} {\bibinfo  {journal} {arXiv
  preprint arXiv:1502.03263}\ } (\bibinfo {year} {2015})}\BibitemShut {NoStop}%
\bibitem [{\citenamefont {Campisi}(2015)}]{Campisi2015a}%
  \BibitemOpen
  \bibfield  {author} {\bibinfo {author} {\bibfnamefont {M.}~\bibnamefont
  {Campisi}},\ }\href {https://doi.org/10.1103/PhysRevE.91.052147} {\bibfield
  {journal} {\bibinfo  {journal} {Phys. Rev. E}\ }\textbf {\bibinfo {volume}
  {91}},\ \bibinfo {pages} {052147} (\bibinfo {year} {2015})}\BibitemShut
  {NoStop}%
\bibitem [{\citenamefont {H\"{a}nggi}\ \emph {et~al.}(2016)\citenamefont
  {H\"{a}nggi}, \citenamefont {Hilbert},\ and\ \citenamefont
  {Dunkel}}]{Hnggi2016}%
  \BibitemOpen
  \bibfield  {author} {\bibinfo {author} {\bibfnamefont {P.}~\bibnamefont
  {H\"{a}nggi}}, \bibinfo {author} {\bibfnamefont {S.}~\bibnamefont
  {Hilbert}},\ and\ \bibinfo {author} {\bibfnamefont {J.}~\bibnamefont
  {Dunkel}},\ }\href {https://doi.org/10.1098/rsta.2015.0039} {\bibfield
  {journal} {\bibinfo  {journal} {Philosophical Transactions of the Royal
  Society A: Mathematical, Physical and Engineering Sciences}\ }\textbf
  {\bibinfo {volume} {374}},\ \bibinfo {pages} {20150039} (\bibinfo {year}
  {2016})}\BibitemShut {NoStop}%
\bibitem [{\citenamefont {Campisi}(2021)}]{campisi2021lectures}%
  \BibitemOpen
  \bibfield  {author} {\bibinfo {author} {\bibfnamefont {M.}~\bibnamefont
  {Campisi}},\ }\href@noop {} {\emph {\bibinfo {title} {Lectures on the
  Mechanical Foundations of Thermodynamics}}}\ (\bibinfo  {publisher}
  {Springer},\ \bibinfo {year} {2021})\BibitemShut {NoStop}%
\bibitem [{\citenamefont {Hartmann}\ \emph {et~al.}(2004)\citenamefont
  {Hartmann}, \citenamefont {Mahler},\ and\ \citenamefont
  {Hess}}]{Hartmann2004}%
  \BibitemOpen
  \bibfield  {author} {\bibinfo {author} {\bibfnamefont {M.}~\bibnamefont
  {Hartmann}}, \bibinfo {author} {\bibfnamefont {G.}~\bibnamefont {Mahler}},\
  and\ \bibinfo {author} {\bibfnamefont {O.}~\bibnamefont {Hess}},\ }\href
  {https://doi.org/10.1103/PhysRevLett.93.080402} {\bibfield  {journal}
  {\bibinfo  {journal} {Phys. Rev. Lett.}\ }\textbf {\bibinfo {volume} {93}},\
  \bibinfo {pages} {080402} (\bibinfo {year} {2004})}\BibitemShut {NoStop}%
\bibitem [{\citenamefont {Ferraro}\ \emph {et~al.}(2012)\citenamefont
  {Ferraro}, \citenamefont {Garc{\'{\i}}a-Saez},\ and\ \citenamefont
  {Ac{\'{\i}}n}}]{Ferraro_2012}%
  \BibitemOpen
  \bibfield  {author} {\bibinfo {author} {\bibfnamefont {A.}~\bibnamefont
  {Ferraro}}, \bibinfo {author} {\bibfnamefont {A.}~\bibnamefont
  {Garc{\'{\i}}a-Saez}},\ and\ \bibinfo {author} {\bibfnamefont
  {A.}~\bibnamefont {Ac{\'{\i}}n}},\ }\href
  {https://doi.org/10.1209/0295-5075/98/10009} {\bibfield  {journal} {\bibinfo
  {journal} {{EPL} (Europhysics Letters)}\ }\textbf {\bibinfo {volume} {98}},\
  \bibinfo {pages} {10009} (\bibinfo {year} {2012})}\BibitemShut {NoStop}%
\bibitem [{\citenamefont {Kliesch}\ \emph {et~al.}(2014)\citenamefont
  {Kliesch}, \citenamefont {Gogolin}, \citenamefont {Kastoryano}, \citenamefont
  {Riera},\ and\ \citenamefont {Eisert}}]{Kliesch2014}%
  \BibitemOpen
  \bibfield  {author} {\bibinfo {author} {\bibfnamefont {M.}~\bibnamefont
  {Kliesch}}, \bibinfo {author} {\bibfnamefont {C.}~\bibnamefont {Gogolin}},
  \bibinfo {author} {\bibfnamefont {M.~J.}\ \bibnamefont {Kastoryano}},
  \bibinfo {author} {\bibfnamefont {A.}~\bibnamefont {Riera}},\ and\ \bibinfo
  {author} {\bibfnamefont {J.}~\bibnamefont {Eisert}},\ }\href
  {https://doi.org/10.1103/PhysRevX.4.031019} {\bibfield  {journal} {\bibinfo
  {journal} {Phys. Rev. X}\ }\textbf {\bibinfo {volume} {4}},\ \bibinfo {pages}
  {031019} (\bibinfo {year} {2014})}\BibitemShut {NoStop}%
\bibitem [{\citenamefont {Hern{\'{a}}ndez-Santana}\ \emph
  {et~al.}(2015)\citenamefont {Hern{\'{a}}ndez-Santana}, \citenamefont {Riera},
  \citenamefont {Hovhannisyan}, \citenamefont {Perarnau-Llobet}, \citenamefont
  {Tagliacozzo},\ and\ \citenamefont {Ac{\'{\i}}n}}]{Hern_ndez_Santana_2015}%
  \BibitemOpen
  \bibfield  {author} {\bibinfo {author} {\bibfnamefont {S.}~\bibnamefont
  {Hern{\'{a}}ndez-Santana}}, \bibinfo {author} {\bibfnamefont
  {A.}~\bibnamefont {Riera}}, \bibinfo {author} {\bibfnamefont {K.~V.}\
  \bibnamefont {Hovhannisyan}}, \bibinfo {author} {\bibfnamefont
  {M.}~\bibnamefont {Perarnau-Llobet}}, \bibinfo {author} {\bibfnamefont
  {L.}~\bibnamefont {Tagliacozzo}},\ and\ \bibinfo {author} {\bibfnamefont
  {A.}~\bibnamefont {Ac{\'{\i}}n}},\ }\href
  {https://doi.org/10.1088/1367-2630/17/8/085007} {\bibfield  {journal}
  {\bibinfo  {journal} {New Journal of Physics}\ }\textbf {\bibinfo {volume}
  {17}},\ \bibinfo {pages} {085007} (\bibinfo {year} {2015})}\BibitemShut
  {NoStop}%
\bibitem [{\citenamefont {Hern{\'{a}}ndez-Santana}\ \emph
  {et~al.}(2021)\citenamefont {Hern{\'{a}}ndez-Santana}, \citenamefont
  {Moln{\'{a}}r}, \citenamefont {Gogolin}, \citenamefont {Cirac},\ and\
  \citenamefont {Ac{\'{\i}}n}}]{Hern_ndez_Santana_2021}%
  \BibitemOpen
  \bibfield  {author} {\bibinfo {author} {\bibfnamefont {S.}~\bibnamefont
  {Hern{\'{a}}ndez-Santana}}, \bibinfo {author} {\bibfnamefont
  {A.}~\bibnamefont {Moln{\'{a}}r}}, \bibinfo {author} {\bibfnamefont
  {C.}~\bibnamefont {Gogolin}}, \bibinfo {author} {\bibfnamefont {J.~I.}\
  \bibnamefont {Cirac}},\ and\ \bibinfo {author} {\bibfnamefont
  {A.}~\bibnamefont {Ac{\'{\i}}n}},\ }\href
  {https://doi.org/10.1088/1367-2630/ac14a9} {\bibfield  {journal} {\bibinfo
  {journal} {New Journal of Physics}\ }\textbf {\bibinfo {volume} {23}},\
  \bibinfo {pages} {073052} (\bibinfo {year} {2021})}\BibitemShut {NoStop}%
\bibitem [{\citenamefont {Casas-V{\'a}zquez}\ and\ \citenamefont
  {Jou}(2003)}]{casas2003temperature}%
  \BibitemOpen
  \bibfield  {author} {\bibinfo {author} {\bibfnamefont {J.}~\bibnamefont
  {Casas-V{\'a}zquez}}\ and\ \bibinfo {author} {\bibfnamefont {D.}~\bibnamefont
  {Jou}},\ }\href@noop {} {\bibfield  {journal} {\bibinfo  {journal} {Reports
  on Progress in Physics}\ }\textbf {\bibinfo {volume} {66}},\ \bibinfo {pages}
  {1937} (\bibinfo {year} {2003})}\BibitemShut {NoStop}%
\bibitem [{\citenamefont {Popov}\ and\ \citenamefont
  {Hernandez}(2007)}]{popov2007ontology}%
  \BibitemOpen
  \bibfield  {author} {\bibinfo {author} {\bibfnamefont {A.~V.}\ \bibnamefont
  {Popov}}\ and\ \bibinfo {author} {\bibfnamefont {R.}~\bibnamefont
  {Hernandez}},\ }\href@noop {} {\bibfield  {journal} {\bibinfo  {journal} {The
  Journal of chemical physics}\ }\textbf {\bibinfo {volume} {126}},\ \bibinfo
  {pages} {244506} (\bibinfo {year} {2007})}\BibitemShut {NoStop}%
\bibitem [{\citenamefont {Puglisi}\ \emph {et~al.}(2017)\citenamefont
  {Puglisi}, \citenamefont {Sarracino},\ and\ \citenamefont
  {Vulpiani}}]{Puglisi2017}%
  \BibitemOpen
  \bibfield  {author} {\bibinfo {author} {\bibfnamefont {A.}~\bibnamefont
  {Puglisi}}, \bibinfo {author} {\bibfnamefont {A.}~\bibnamefont {Sarracino}},\
  and\ \bibinfo {author} {\bibfnamefont {A.}~\bibnamefont {Vulpiani}},\ }\href
  {https://doi.org/10.1016/j.physrep.2017.09.001} {\bibfield  {journal}
  {\bibinfo  {journal} {Physics Reports}\ }\textbf {\bibinfo {volume}
  {709-710}},\ \bibinfo {pages} {1} (\bibinfo {year} {2017})}\BibitemShut
  {NoStop}%
\bibitem [{\citenamefont {Hsiang}\ and\ \citenamefont
  {Hu}(2021)}]{Hsiang_2021}%
  \BibitemOpen
  \bibfield  {author} {\bibinfo {author} {\bibfnamefont {J.-T.}\ \bibnamefont
  {Hsiang}}\ and\ \bibinfo {author} {\bibfnamefont {B.-L.}\ \bibnamefont
  {Hu}},\ }\bibfield  {journal} {\bibinfo  {journal} {Physical Review D}\
  }\textbf {\bibinfo {volume} {103}},\ \href
  {https://doi.org/10.1103/physrevd.103.065001} {10.1103/physrevd.103.065001}
  (\bibinfo {year} {2021})\BibitemShut {NoStop}%
\bibitem [{Note1()}]{Note1}%
  \BibitemOpen
  \bibinfo {note} {More precisely, they become indistinguishable for physically
  relevant local observables and under mild conditions for $\rho $ and $H$, see
  e.g. counterexamples in integrable systems \cite
  {kinoshita2006quantum,Brenes2020} and many-body localisation \cite
  {anderson1958absence,Abanin_2019,nandkishore2015many}}\BibitemShut {NoStop}%
\bibitem [{\citenamefont {Deutsch}(1991)}]{Deutsch1991}%
  \BibitemOpen
  \bibfield  {author} {\bibinfo {author} {\bibfnamefont {J.~M.}\ \bibnamefont
  {Deutsch}},\ }\href {https://doi.org/10.1103/PhysRevA.43.2046} {\bibfield
  {journal} {\bibinfo  {journal} {Phys. Rev. A}\ }\textbf {\bibinfo {volume}
  {43}},\ \bibinfo {pages} {2046} (\bibinfo {year} {1991})}\BibitemShut
  {NoStop}%
\bibitem [{\citenamefont {Srednicki}(1994)}]{Srednicki1994}%
  \BibitemOpen
  \bibfield  {author} {\bibinfo {author} {\bibfnamefont {M.}~\bibnamefont
  {Srednicki}},\ }\href {https://doi.org/10.1103/PhysRevE.50.888} {\bibfield
  {journal} {\bibinfo  {journal} {Phys. Rev. E}\ }\textbf {\bibinfo {volume}
  {50}},\ \bibinfo {pages} {888} (\bibinfo {year} {1994})}\BibitemShut
  {NoStop}%
\bibitem [{\citenamefont {Rigol}\ \emph {et~al.}(2008)\citenamefont {Rigol},
  \citenamefont {Dunjko},\ and\ \citenamefont {Olshanii}}]{Rigol2008}%
  \BibitemOpen
  \bibfield  {author} {\bibinfo {author} {\bibfnamefont {M.}~\bibnamefont
  {Rigol}}, \bibinfo {author} {\bibfnamefont {V.}~\bibnamefont {Dunjko}},\ and\
  \bibinfo {author} {\bibfnamefont {M.}~\bibnamefont {Olshanii}},\ }\href
  {https://doi.org/10.1038/nature06838} {\bibfield  {journal} {\bibinfo
  {journal} {Nature}\ }\textbf {\bibinfo {volume} {452}},\ \bibinfo {pages}
  {854} (\bibinfo {year} {2008})}\BibitemShut {NoStop}%
\bibitem [{\citenamefont {Polkovnikov}\ \emph {et~al.}(2011)\citenamefont
  {Polkovnikov}, \citenamefont {Sengupta}, \citenamefont {Silva},\ and\
  \citenamefont {Vengalattore}}]{Polkovnikov2011}%
  \BibitemOpen
  \bibfield  {author} {\bibinfo {author} {\bibfnamefont {A.}~\bibnamefont
  {Polkovnikov}}, \bibinfo {author} {\bibfnamefont {K.}~\bibnamefont
  {Sengupta}}, \bibinfo {author} {\bibfnamefont {A.}~\bibnamefont {Silva}},\
  and\ \bibinfo {author} {\bibfnamefont {M.}~\bibnamefont {Vengalattore}},\
  }\href {https://doi.org/10.1103/RevModPhys.83.863} {\bibfield  {journal}
  {\bibinfo  {journal} {Rev. Mod. Phys.}\ }\textbf {\bibinfo {volume} {83}},\
  \bibinfo {pages} {863} (\bibinfo {year} {2011})}\BibitemShut {NoStop}%
\bibitem [{\citenamefont {Eisert}\ \emph {et~al.}(2015)\citenamefont {Eisert},
  \citenamefont {Friesdorf},\ and\ \citenamefont {Gogolin}}]{Eisert2015}%
  \BibitemOpen
  \bibfield  {author} {\bibinfo {author} {\bibfnamefont {J.}~\bibnamefont
  {Eisert}}, \bibinfo {author} {\bibfnamefont {M.}~\bibnamefont {Friesdorf}},\
  and\ \bibinfo {author} {\bibfnamefont {C.}~\bibnamefont {Gogolin}},\ }\href
  {https://doi.org/10.1038/nphys3215} {\bibfield  {journal} {\bibinfo
  {journal} {Nature Physics}\ }\textbf {\bibinfo {volume} {11}},\ \bibinfo
  {pages} {124} (\bibinfo {year} {2015})}\BibitemShut {NoStop}%
\bibitem [{\citenamefont {Gogolin}\ and\ \citenamefont
  {Eisert}(2016)}]{Gogolin2016}%
  \BibitemOpen
  \bibfield  {author} {\bibinfo {author} {\bibfnamefont {C.}~\bibnamefont
  {Gogolin}}\ and\ \bibinfo {author} {\bibfnamefont {J.}~\bibnamefont
  {Eisert}},\ }\href {https://doi.org/10.1088/0034-4885/79/5/056001} {\bibfield
   {journal} {\bibinfo  {journal} {Reports on Progress in Physics}\ }\textbf
  {\bibinfo {volume} {79}},\ \bibinfo {pages} {056001} (\bibinfo {year}
  {2016})}\BibitemShut {NoStop}%
\bibitem [{\citenamefont {Mitchison}\ \emph {et~al.}(2022)\citenamefont
  {Mitchison}, \citenamefont {Purkayastha}, \citenamefont {Brenes},
  \citenamefont {Silva},\ and\ \citenamefont {Goold}}]{Mitchison2022}%
  \BibitemOpen
  \bibfield  {author} {\bibinfo {author} {\bibfnamefont {M.~T.}\ \bibnamefont
  {Mitchison}}, \bibinfo {author} {\bibfnamefont {A.}~\bibnamefont
  {Purkayastha}}, \bibinfo {author} {\bibfnamefont {M.}~\bibnamefont {Brenes}},
  \bibinfo {author} {\bibfnamefont {A.}~\bibnamefont {Silva}},\ and\ \bibinfo
  {author} {\bibfnamefont {J.}~\bibnamefont {Goold}},\ }\href
  {https://doi.org/10.1103/PhysRevA.105.L030201} {\bibfield  {journal}
  {\bibinfo  {journal} {Phys. Rev. A}\ }\textbf {\bibinfo {volume} {105}},\
  \bibinfo {pages} {L030201} (\bibinfo {year} {2022})}\BibitemShut {NoStop}%
\bibitem [{\citenamefont {Schnack}(1998)}]{Schnack1998}%
  \BibitemOpen
  \bibfield  {author} {\bibinfo {author} {\bibfnamefont {J.}~\bibnamefont
  {Schnack}},\ }\href
  {https://doi.org/https://doi.org/10.1016/S0378-4371(98)00236-2} {\bibfield
  {journal} {\bibinfo  {journal} {Physica A: Statistical Mechanics and its
  Applications}\ }\textbf {\bibinfo {volume} {259}},\ \bibinfo {pages} {49}
  (\bibinfo {year} {1998})}\BibitemShut {NoStop}%
\bibitem [{\citenamefont {Alipour}\ \emph {et~al.}(2021)\citenamefont
  {Alipour}, \citenamefont {Benatti}, \citenamefont {Afsary}, \citenamefont
  {Bakhshinezhad}, \citenamefont {Ramezani}, \citenamefont {Ala-Nissila},\ and\
  \citenamefont {Rezakhani}}]{Alipour2021}%
  \BibitemOpen
  \bibfield  {author} {\bibinfo {author} {\bibfnamefont {S.}~\bibnamefont
  {Alipour}}, \bibinfo {author} {\bibfnamefont {F.}~\bibnamefont {Benatti}},
  \bibinfo {author} {\bibfnamefont {M.}~\bibnamefont {Afsary}}, \bibinfo
  {author} {\bibfnamefont {F.}~\bibnamefont {Bakhshinezhad}}, \bibinfo {author}
  {\bibfnamefont {M.}~\bibnamefont {Ramezani}}, \bibinfo {author}
  {\bibfnamefont {T.}~\bibnamefont {Ala-Nissila}},\ and\ \bibinfo {author}
  {\bibfnamefont {A.~T.}\ \bibnamefont {Rezakhani}},\ }\href
  {https://doi.org/10.48550/ARXIV.2105.11915} {\bibinfo {title} {Temperature in
  nonequilibrium quantum systems}} (\bibinfo {year} {2021})\BibitemShut
  {NoStop}%
\bibitem [{\citenamefont {Brunner}\ \emph {et~al.}(2012)\citenamefont
  {Brunner}, \citenamefont {Linden}, \citenamefont {Popescu},\ and\
  \citenamefont {Skrzypczyk}}]{Brunner2012}%
  \BibitemOpen
  \bibfield  {author} {\bibinfo {author} {\bibfnamefont {N.}~\bibnamefont
  {Brunner}}, \bibinfo {author} {\bibfnamefont {N.}~\bibnamefont {Linden}},
  \bibinfo {author} {\bibfnamefont {S.}~\bibnamefont {Popescu}},\ and\ \bibinfo
  {author} {\bibfnamefont {P.}~\bibnamefont {Skrzypczyk}},\ }\bibfield
  {journal} {\bibinfo  {journal} {Physical Review E}\ }\textbf {\bibinfo
  {volume} {85}},\ \href {https://doi.org/10.1103/physreve.85.051117}
  {10.1103/physreve.85.051117} (\bibinfo {year} {2012})\BibitemShut {NoStop}%
\bibitem [{\citenamefont {Skrzypczyk}\ \emph {et~al.}(2015)\citenamefont
  {Skrzypczyk}, \citenamefont {Silva},\ and\ \citenamefont
  {Brunner}}]{skrzypczyk2015passivity}%
  \BibitemOpen
  \bibfield  {author} {\bibinfo {author} {\bibfnamefont {P.}~\bibnamefont
  {Skrzypczyk}}, \bibinfo {author} {\bibfnamefont {R.}~\bibnamefont {Silva}},\
  and\ \bibinfo {author} {\bibfnamefont {N.}~\bibnamefont {Brunner}},\ }\href
  {https://doi.org/10.1103/PhysRevE.91.052133} {\bibfield  {journal} {\bibinfo
  {journal} {Phys. Rev. E}\ }\textbf {\bibinfo {volume} {91}},\ \bibinfo
  {pages} {052133} (\bibinfo {year} {2015})}\BibitemShut {NoStop}%
\bibitem [{\citenamefont {Silva}\ \emph {et~al.}(2016)\citenamefont {Silva},
  \citenamefont {Manzano}, \citenamefont {Skrzypczyk},\ and\ \citenamefont
  {Brunner}}]{Silva2016}%
  \BibitemOpen
  \bibfield  {author} {\bibinfo {author} {\bibfnamefont {R.}~\bibnamefont
  {Silva}}, \bibinfo {author} {\bibfnamefont {G.}~\bibnamefont {Manzano}},
  \bibinfo {author} {\bibfnamefont {P.}~\bibnamefont {Skrzypczyk}},\ and\
  \bibinfo {author} {\bibfnamefont {N.}~\bibnamefont {Brunner}},\ }\href
  {https://doi.org/10.1103/PhysRevE.94.032120} {\bibfield  {journal} {\bibinfo
  {journal} {Phys. Rev. E}\ }\textbf {\bibinfo {volume} {94}},\ \bibinfo
  {pages} {032120} (\bibinfo {year} {2016})}\BibitemShut {NoStop}%
\bibitem [{\citenamefont {Mitchison}(2019)}]{Mitchison2019}%
  \BibitemOpen
  \bibfield  {author} {\bibinfo {author} {\bibfnamefont {M.~T.}\ \bibnamefont
  {Mitchison}},\ }\href {https://doi.org/10.1080/00107514.2019.1631555}
  {\bibfield  {journal} {\bibinfo  {journal} {Contemporary Physics}\ }\textbf
  {\bibinfo {volume} {60}},\ \bibinfo {pages} {164–187} (\bibinfo {year}
  {2019})}\BibitemShut {NoStop}%
\bibitem [{\citenamefont {Ro\ss{}nagel}\ \emph {et~al.}(2014)\citenamefont
  {Ro\ss{}nagel}, \citenamefont {Abah}, \citenamefont {Schmidt-Kaler},
  \citenamefont {Singer},\ and\ \citenamefont {Lutz}}]{Rossnagel2014}%
  \BibitemOpen
  \bibfield  {author} {\bibinfo {author} {\bibfnamefont {J.}~\bibnamefont
  {Ro\ss{}nagel}}, \bibinfo {author} {\bibfnamefont {O.}~\bibnamefont {Abah}},
  \bibinfo {author} {\bibfnamefont {F.}~\bibnamefont {Schmidt-Kaler}}, \bibinfo
  {author} {\bibfnamefont {K.}~\bibnamefont {Singer}},\ and\ \bibinfo {author}
  {\bibfnamefont {E.}~\bibnamefont {Lutz}},\ }\href
  {https://doi.org/10.1103/PhysRevLett.112.030602} {\bibfield  {journal}
  {\bibinfo  {journal} {Phys. Rev. Lett.}\ }\textbf {\bibinfo {volume} {112}},\
  \bibinfo {pages} {030602} (\bibinfo {year} {2014})}\BibitemShut {NoStop}%
\bibitem [{\citenamefont {Abah}\ and\ \citenamefont {Lutz}(2014)}]{Abah_2014}%
  \BibitemOpen
  \bibfield  {author} {\bibinfo {author} {\bibfnamefont {O.}~\bibnamefont
  {Abah}}\ and\ \bibinfo {author} {\bibfnamefont {E.}~\bibnamefont {Lutz}},\
  }\href {https://doi.org/10.1209/0295-5075/106/20001} {\bibfield  {journal}
  {\bibinfo  {journal} {{EPL} (Europhysics Letters)}\ }\textbf {\bibinfo
  {volume} {106}},\ \bibinfo {pages} {20001} (\bibinfo {year}
  {2014})}\BibitemShut {NoStop}%
\bibitem [{\citenamefont {Correa}\ \emph {et~al.}(2014)\citenamefont {Correa},
  \citenamefont {Palao}, \citenamefont {Alonso},\ and\ \citenamefont
  {Adesso}}]{correa2014quantum}%
  \BibitemOpen
  \bibfield  {author} {\bibinfo {author} {\bibfnamefont {L.~A.}\ \bibnamefont
  {Correa}}, \bibinfo {author} {\bibfnamefont {J.~P.}\ \bibnamefont {Palao}},
  \bibinfo {author} {\bibfnamefont {D.}~\bibnamefont {Alonso}},\ and\ \bibinfo
  {author} {\bibfnamefont {G.}~\bibnamefont {Adesso}},\ }\href
  {https://doi.org/10.1038/srep03949} {\bibfield  {journal} {\bibinfo
  {journal} {Scientific reports}\ }\textbf {\bibinfo {volume} {4}},\ \bibinfo
  {pages} {1} (\bibinfo {year} {2014})}\BibitemShut {NoStop}%
\bibitem [{\citenamefont {Janzing}\ \emph {et~al.}(2000)\citenamefont
  {Janzing}, \citenamefont {Wocjan}, \citenamefont {Zeier}, \citenamefont
  {Geiss},\ and\ \citenamefont {Beth}}]{Janzing2000}%
  \BibitemOpen
  \bibfield  {author} {\bibinfo {author} {\bibfnamefont {D.}~\bibnamefont
  {Janzing}}, \bibinfo {author} {\bibfnamefont {P.}~\bibnamefont {Wocjan}},
  \bibinfo {author} {\bibfnamefont {R.}~\bibnamefont {Zeier}}, \bibinfo
  {author} {\bibfnamefont {R.}~\bibnamefont {Geiss}},\ and\ \bibinfo {author}
  {\bibfnamefont {T.}~\bibnamefont {Beth}},\ }\href
  {https://doi.org/10.1023/A:1026422630734} {\bibfield  {journal} {\bibinfo
  {journal} {Int. J. Theor. Phys.}\ }\textbf {\bibinfo {volume} {39}},\
  \bibinfo {pages} {2717} (\bibinfo {year} {2000})}\BibitemShut {NoStop}%
\bibitem [{\citenamefont {Quan}\ \emph {et~al.}(2007)\citenamefont {Quan},
  \citenamefont {Liu}, \citenamefont {Sun},\ and\ \citenamefont
  {Nori}}]{Quan2007}%
  \BibitemOpen
  \bibfield  {author} {\bibinfo {author} {\bibfnamefont {H.~T.}\ \bibnamefont
  {Quan}}, \bibinfo {author} {\bibfnamefont {Y.-x.}\ \bibnamefont {Liu}},
  \bibinfo {author} {\bibfnamefont {C.~P.}\ \bibnamefont {Sun}},\ and\ \bibinfo
  {author} {\bibfnamefont {F.}~\bibnamefont {Nori}},\ }\href
  {https://doi.org/10.1103/PhysRevE.76.031105} {\bibfield  {journal} {\bibinfo
  {journal} {Phys. Rev. E}\ }\textbf {\bibinfo {volume} {76}},\ \bibinfo
  {pages} {031105} (\bibinfo {year} {2007})}\BibitemShut {NoStop}%
\bibitem [{\citenamefont {Koukoulekidis}\ \emph {et~al.}(2021)\citenamefont
  {Koukoulekidis}, \citenamefont {Alexander}, \citenamefont {Hebdige},\ and\
  \citenamefont {Jennings}}]{Koukoulekidis2021}%
  \BibitemOpen
  \bibfield  {author} {\bibinfo {author} {\bibfnamefont {N.}~\bibnamefont
  {Koukoulekidis}}, \bibinfo {author} {\bibfnamefont {R.}~\bibnamefont
  {Alexander}}, \bibinfo {author} {\bibfnamefont {T.}~\bibnamefont {Hebdige}},\
  and\ \bibinfo {author} {\bibfnamefont {D.}~\bibnamefont {Jennings}},\ }\href
  {https://doi.org/10.22331/q-2021-03-15-411} {\bibfield  {journal} {\bibinfo
  {journal} {Quantum}\ }\textbf {\bibinfo {volume} {5}},\ \bibinfo {pages}
  {411} (\bibinfo {year} {2021})}\BibitemShut {NoStop}%
\bibitem [{\citenamefont {Jonathan}\ and\ \citenamefont
  {Plenio}(1999)}]{Jonathan_1999}%
  \BibitemOpen
  \bibfield  {author} {\bibinfo {author} {\bibfnamefont {D.}~\bibnamefont
  {Jonathan}}\ and\ \bibinfo {author} {\bibfnamefont {M.~B.}\ \bibnamefont
  {Plenio}},\ }\href {https://doi.org/10.1103/physrevlett.83.3566} {\bibfield
  {journal} {\bibinfo  {journal} {Phys. Rev. Lett.}\ }\textbf {\bibinfo
  {volume} {83}},\ \bibinfo {pages} {3566–3569} (\bibinfo {year}
  {1999})}\BibitemShut {NoStop}%
\bibitem [{\citenamefont {Brandão}\ \emph {et~al.}(2015)\citenamefont
  {Brandão}, \citenamefont {Horodecki}, \citenamefont {Ng}, \citenamefont
  {Oppenheim},\ and\ \citenamefont {Wehner}}]{Brand_o_2015}%
  \BibitemOpen
  \bibfield  {author} {\bibinfo {author} {\bibfnamefont {F.}~\bibnamefont
  {Brandão}}, \bibinfo {author} {\bibfnamefont {M.}~\bibnamefont {Horodecki}},
  \bibinfo {author} {\bibfnamefont {N.}~\bibnamefont {Ng}}, \bibinfo {author}
  {\bibfnamefont {J.}~\bibnamefont {Oppenheim}},\ and\ \bibinfo {author}
  {\bibfnamefont {S.}~\bibnamefont {Wehner}},\ }\href
  {https://doi.org/10.1073/pnas.1411728112} {\bibfield  {journal} {\bibinfo
  {journal} {PNAS}\ }\textbf {\bibinfo {volume} {112}},\ \bibinfo {pages}
  {3275–3279} (\bibinfo {year} {2015})}\BibitemShut {NoStop}%
\bibitem [{\citenamefont {M\"uller}(2018)}]{Mueller2018}%
  \BibitemOpen
  \bibfield  {author} {\bibinfo {author} {\bibfnamefont {M.~P.}\ \bibnamefont
  {M\"uller}},\ }\href {https://doi.org/10.1103/PhysRevX.8.041051} {\bibfield
  {journal} {\bibinfo  {journal} {Phys. Rev. X}\ }\textbf {\bibinfo {volume}
  {8}},\ \bibinfo {pages} {041051} (\bibinfo {year} {2018})}\BibitemShut
  {NoStop}%
\bibitem [{\citenamefont {Ng}\ \emph {et~al.}(2015)\citenamefont {Ng},
  \citenamefont {Mančinska}, \citenamefont {Cirstoiu}, \citenamefont
  {Eisert},\ and\ \citenamefont {Wehner}}]{Ng_2015}%
  \BibitemOpen
  \bibfield  {author} {\bibinfo {author} {\bibfnamefont {N.~H.~Y.}\
  \bibnamefont {Ng}}, \bibinfo {author} {\bibfnamefont {L.}~\bibnamefont
  {Mančinska}}, \bibinfo {author} {\bibfnamefont {C.}~\bibnamefont
  {Cirstoiu}}, \bibinfo {author} {\bibfnamefont {J.}~\bibnamefont {Eisert}},\
  and\ \bibinfo {author} {\bibfnamefont {S.}~\bibnamefont {Wehner}},\ }\href
  {https://doi.org/10.1088/1367-2630/17/8/085004} {\bibfield  {journal}
  {\bibinfo  {journal} {New J. Phys.}\ }\textbf {\bibinfo {volume} {17}},\
  \bibinfo {pages} {085004} (\bibinfo {year} {2015})}\BibitemShut {NoStop}%
\bibitem [{\citenamefont {Wilming}\ and\ \citenamefont
  {Gallego}(2017)}]{Wilming2017}%
  \BibitemOpen
  \bibfield  {author} {\bibinfo {author} {\bibfnamefont {H.}~\bibnamefont
  {Wilming}}\ and\ \bibinfo {author} {\bibfnamefont {R.}~\bibnamefont
  {Gallego}},\ }\href {https://doi.org/10.1103/PhysRevX.7.041033} {\bibfield
  {journal} {\bibinfo  {journal} {Phys. Rev. X}\ }\textbf {\bibinfo {volume}
  {7}},\ \bibinfo {pages} {041033} (\bibinfo {year} {2017})}\BibitemShut
  {NoStop}%
\bibitem [{\citenamefont {Shiraishi}\ and\ \citenamefont
  {Sagawa}(2021)}]{shiraishi2021quantum}%
  \BibitemOpen
  \bibfield  {author} {\bibinfo {author} {\bibfnamefont {N.}~\bibnamefont
  {Shiraishi}}\ and\ \bibinfo {author} {\bibfnamefont {T.}~\bibnamefont
  {Sagawa}},\ }\href@noop {} {\bibfield  {journal} {\bibinfo  {journal}
  {Physical Review Letters}\ }\textbf {\bibinfo {volume} {126}},\ \bibinfo
  {pages} {150502} (\bibinfo {year} {2021})}\BibitemShut {NoStop}%
\bibitem [{\citenamefont {Wilming}(2021)}]{wilming2021entropy}%
  \BibitemOpen
  \bibfield  {author} {\bibinfo {author} {\bibfnamefont {H.}~\bibnamefont
  {Wilming}},\ }\href@noop {} {\bibfield  {journal} {\bibinfo  {journal}
  {Physical review letters}\ }\textbf {\bibinfo {volume} {127}},\ \bibinfo
  {pages} {260402} (\bibinfo {year} {2021})}\BibitemShut {NoStop}%
\bibitem [{\citenamefont {Korzekwa}\ and\ \citenamefont
  {Lostaglio}(2022)}]{Korzekwa2022}%
  \BibitemOpen
  \bibfield  {author} {\bibinfo {author} {\bibfnamefont {K.}~\bibnamefont
  {Korzekwa}}\ and\ \bibinfo {author} {\bibfnamefont {M.}~\bibnamefont
  {Lostaglio}},\ }\href {https://doi.org/10.48550/ARXIV.2202.12616} {\bibinfo
  {title} {Optimizing thermalizations}} (\bibinfo {year} {2022})\BibitemShut
  {NoStop}%
\bibitem [{\citenamefont {Callen}(1998)}]{callen1998thermodynamics}%
  \BibitemOpen
  \bibfield  {author} {\bibinfo {author} {\bibfnamefont {H.~B.}\ \bibnamefont
  {Callen}},\ }\href@noop {} {\emph {\bibinfo {title} {Thermodynamics and an
  Introduction to Thermostatistics}}}\ (\bibinfo  {publisher} {American
  Association of Physics Teachers},\ \bibinfo {year} {1998})\BibitemShut
  {NoStop}%
\bibitem [{\citenamefont {Lieb}\ and\ \citenamefont
  {Yngvason}(1999)}]{lieb1999physics}%
  \BibitemOpen
  \bibfield  {author} {\bibinfo {author} {\bibfnamefont {E.~H.}\ \bibnamefont
  {Lieb}}\ and\ \bibinfo {author} {\bibfnamefont {J.}~\bibnamefont
  {Yngvason}},\ }\href@noop {} {\bibfield  {journal} {\bibinfo  {journal}
  {Physics Reports}\ }\textbf {\bibinfo {volume} {310}},\ \bibinfo {pages} {1}
  (\bibinfo {year} {1999})}\BibitemShut {NoStop}%
\bibitem [{\citenamefont {D'Alessandro}(2007)}]{DAlessandro2007}%
  \BibitemOpen
  \bibfield  {author} {\bibinfo {author} {\bibfnamefont {D.}~\bibnamefont
  {D'Alessandro}},\ }\href {https://books.google.ch/books?id=e5M0id5enzQC}
  {\emph {\bibinfo {title} {Introduction to Quantum Control and Dynamics}}},\
  Chapman \& Hall/CRC Applied Mathematics \& Nonlinear Science\ (\bibinfo
  {publisher} {CRC Press},\ \bibinfo {year} {2007})\BibitemShut {NoStop}%
\bibitem [{\citenamefont {Lloyd}(1989)}]{Loyd19898}%
  \BibitemOpen
  \bibfield  {author} {\bibinfo {author} {\bibfnamefont {S.}~\bibnamefont
  {Lloyd}},\ }\href {https://doi.org/10.1103/PhysRevA.39.5378} {\bibfield
  {journal} {\bibinfo  {journal} {Phys. Rev. A}\ }\textbf {\bibinfo {volume}
  {39}},\ \bibinfo {pages} {5378} (\bibinfo {year} {1989})}\BibitemShut
  {NoStop}%
\bibitem [{\citenamefont {Jennings}\ and\ \citenamefont
  {Rudolph}(2010)}]{jennings2010entanglement}%
  \BibitemOpen
  \bibfield  {author} {\bibinfo {author} {\bibfnamefont {D.}~\bibnamefont
  {Jennings}}\ and\ \bibinfo {author} {\bibfnamefont {T.}~\bibnamefont
  {Rudolph}},\ }\href@noop {} {\bibfield  {journal} {\bibinfo  {journal}
  {Physical Review E}\ }\textbf {\bibinfo {volume} {81}},\ \bibinfo {pages}
  {061130} (\bibinfo {year} {2010})}\BibitemShut {NoStop}%
\bibitem [{\citenamefont {Micadei}\ \emph {et~al.}(2019)\citenamefont
  {Micadei}, \citenamefont {Peterson}, \citenamefont {Souza}, \citenamefont
  {Sarthour}, \citenamefont {Oliveira}, \citenamefont {Landi}, \citenamefont
  {Batalh{\~a}o}, \citenamefont {Serra},\ and\ \citenamefont
  {Lutz}}]{micadei2019reversing}%
  \BibitemOpen
  \bibfield  {author} {\bibinfo {author} {\bibfnamefont {K.}~\bibnamefont
  {Micadei}}, \bibinfo {author} {\bibfnamefont {J.~P.}\ \bibnamefont
  {Peterson}}, \bibinfo {author} {\bibfnamefont {A.~M.}\ \bibnamefont {Souza}},
  \bibinfo {author} {\bibfnamefont {R.~S.}\ \bibnamefont {Sarthour}}, \bibinfo
  {author} {\bibfnamefont {I.~S.}\ \bibnamefont {Oliveira}}, \bibinfo {author}
  {\bibfnamefont {G.~T.}\ \bibnamefont {Landi}}, \bibinfo {author}
  {\bibfnamefont {T.~B.}\ \bibnamefont {Batalh{\~a}o}}, \bibinfo {author}
  {\bibfnamefont {R.~M.}\ \bibnamefont {Serra}},\ and\ \bibinfo {author}
  {\bibfnamefont {E.}~\bibnamefont {Lutz}},\ }\href@noop {} {\bibfield
  {journal} {\bibinfo  {journal} {Nature communications}\ }\textbf {\bibinfo
  {volume} {10}},\ \bibinfo {pages} {1} (\bibinfo {year} {2019})}\BibitemShut
  {NoStop}%
\bibitem [{\citenamefont {Henao}\ and\ \citenamefont
  {Serra}(2018)}]{Henao_2018}%
  \BibitemOpen
  \bibfield  {author} {\bibinfo {author} {\bibfnamefont {I.}~\bibnamefont
  {Henao}}\ and\ \bibinfo {author} {\bibfnamefont {R.~M.}\ \bibnamefont
  {Serra}},\ }\bibfield  {journal} {\bibinfo  {journal} {Physical Review E}\
  }\textbf {\bibinfo {volume} {97}},\ \href
  {https://doi.org/10.1103/physreve.97.062105} {10.1103/physreve.97.062105}
  (\bibinfo {year} {2018})\BibitemShut {NoStop}%
\bibitem [{\citenamefont {Horodecki}\ and\ \citenamefont
  {Oppenheim}(2013)}]{Horodecki2013}%
  \BibitemOpen
  \bibfield  {author} {\bibinfo {author} {\bibfnamefont {M.}~\bibnamefont
  {Horodecki}}\ and\ \bibinfo {author} {\bibfnamefont {J.}~\bibnamefont
  {Oppenheim}},\ }\href {https://doi.org/10.1038/ncomms3059} {\bibfield
  {journal} {\bibinfo  {journal} {Nat. Commun.}\ }\textbf {\bibinfo {volume}
  {4}},\ \bibinfo {pages} {2059} (\bibinfo {year} {2013})}\BibitemShut
  {NoStop}%
\bibitem [{\citenamefont {Lostaglio}(2019)}]{Lostaglio2019}%
  \BibitemOpen
  \bibfield  {author} {\bibinfo {author} {\bibfnamefont {M.}~\bibnamefont
  {Lostaglio}},\ }\href {https://doi.org/10.1088/1361-6633/ab46e5} {\bibfield
  {journal} {\bibinfo  {journal} {Rep. Prog. Phys.}\ }\textbf {\bibinfo
  {volume} {82}},\ \bibinfo {pages} {114001} (\bibinfo {year}
  {2019})}\BibitemShut {NoStop}%
\bibitem [{\citenamefont {Korzekwa}\ \emph {et~al.}(2016)\citenamefont
  {Korzekwa}, \citenamefont {Lostaglio}, \citenamefont {Oppenheim},\ and\
  \citenamefont {Jennings}}]{Korzekwa2016}%
  \BibitemOpen
  \bibfield  {author} {\bibinfo {author} {\bibfnamefont {K.}~\bibnamefont
  {Korzekwa}}, \bibinfo {author} {\bibfnamefont {M.}~\bibnamefont {Lostaglio}},
  \bibinfo {author} {\bibfnamefont {J.}~\bibnamefont {Oppenheim}},\ and\
  \bibinfo {author} {\bibfnamefont {D.}~\bibnamefont {Jennings}},\ }\href
  {https://doi.org/10.1088/1367-2630/18/2/023045} {\bibfield  {journal}
  {\bibinfo  {journal} {New J. of Phys.}\ }\textbf {\bibinfo {volume} {18}},\
  \bibinfo {pages} {023045} (\bibinfo {year} {2016})}\BibitemShut {NoStop}%
\bibitem [{\citenamefont {Chubb}\ \emph {et~al.}(2018)\citenamefont {Chubb},
  \citenamefont {Tomamichel},\ and\ \citenamefont {Korzekwa}}]{Chubb_2018}%
  \BibitemOpen
  \bibfield  {author} {\bibinfo {author} {\bibfnamefont {C.~T.}\ \bibnamefont
  {Chubb}}, \bibinfo {author} {\bibfnamefont {M.}~\bibnamefont {Tomamichel}},\
  and\ \bibinfo {author} {\bibfnamefont {K.}~\bibnamefont {Korzekwa}},\ }\href
  {https://doi.org/10.22331/q-2018-11-27-108} {\bibfield  {journal} {\bibinfo
  {journal} {Quantum}\ }\textbf {\bibinfo {volume} {2}},\ \bibinfo {pages}
  {108} (\bibinfo {year} {2018})}\BibitemShut {NoStop}%
\bibitem [{\citenamefont {Lostaglio}\ \emph
  {et~al.}(2015{\natexlab{a}})\citenamefont {Lostaglio}, \citenamefont
  {Jennings},\ and\ \citenamefont {Rudolph}}]{lostaglio2015description}%
  \BibitemOpen
  \bibfield  {author} {\bibinfo {author} {\bibfnamefont {M.}~\bibnamefont
  {Lostaglio}}, \bibinfo {author} {\bibfnamefont {D.}~\bibnamefont
  {Jennings}},\ and\ \bibinfo {author} {\bibfnamefont {T.}~\bibnamefont
  {Rudolph}},\ }\href@noop {} {\bibfield  {journal} {\bibinfo  {journal} {Nat.
  Commun.}\ }\textbf {\bibinfo {volume} {6}},\ \bibinfo {pages} {1} (\bibinfo
  {year} {2015}{\natexlab{a}})}\BibitemShut {NoStop}%
\bibitem [{\citenamefont {Lostaglio}\ \emph
  {et~al.}(2015{\natexlab{b}})\citenamefont {Lostaglio}, \citenamefont
  {Korzekwa}, \citenamefont {Jennings},\ and\ \citenamefont
  {Rudolph}}]{lostaglio2015quantum}%
  \BibitemOpen
  \bibfield  {author} {\bibinfo {author} {\bibfnamefont {M.}~\bibnamefont
  {Lostaglio}}, \bibinfo {author} {\bibfnamefont {K.}~\bibnamefont {Korzekwa}},
  \bibinfo {author} {\bibfnamefont {D.}~\bibnamefont {Jennings}},\ and\
  \bibinfo {author} {\bibfnamefont {T.}~\bibnamefont {Rudolph}},\ }\href@noop
  {} {\bibfield  {journal} {\bibinfo  {journal} {Physical review X}\ }\textbf
  {\bibinfo {volume} {5}},\ \bibinfo {pages} {021001} (\bibinfo {year}
  {2015}{\natexlab{b}})}\BibitemShut {NoStop}%
\bibitem [{\citenamefont {Marvian}\ and\ \citenamefont
  {Spekkens}(2019)}]{Marvian2019}%
  \BibitemOpen
  \bibfield  {author} {\bibinfo {author} {\bibfnamefont {I.}~\bibnamefont
  {Marvian}}\ and\ \bibinfo {author} {\bibfnamefont {R.~W.}\ \bibnamefont
  {Spekkens}},\ }\href {https://doi.org/10.1103/PhysRevLett.123.020404}
  {\bibfield  {journal} {\bibinfo  {journal} {Phys. Rev. Lett.}\ }\textbf
  {\bibinfo {volume} {123}},\ \bibinfo {pages} {020404} (\bibinfo {year}
  {2019})}\BibitemShut {NoStop}%
\bibitem [{\citenamefont {Alicki}\ and\ \citenamefont
  {Fannes}(2013)}]{Alicki2013}%
  \BibitemOpen
  \bibfield  {author} {\bibinfo {author} {\bibfnamefont {R.}~\bibnamefont
  {Alicki}}\ and\ \bibinfo {author} {\bibfnamefont {M.}~\bibnamefont
  {Fannes}},\ }\href {https://doi.org/10.1103/PhysRevE.87.042123} {\bibfield
  {journal} {\bibinfo  {journal} {Phys. Rev. E}\ }\textbf {\bibinfo {volume}
  {87}},\ \bibinfo {pages} {042123} (\bibinfo {year} {2013})}\BibitemShut
  {NoStop}%
\bibitem [{\citenamefont {Salvia}\ and\ \citenamefont
  {Giovannetti}(2020)}]{salvia2020energy}%
  \BibitemOpen
  \bibfield  {author} {\bibinfo {author} {\bibfnamefont {R.}~\bibnamefont
  {Salvia}}\ and\ \bibinfo {author} {\bibfnamefont {V.}~\bibnamefont
  {Giovannetti}},\ }\href@noop {} {\bibfield  {journal} {\bibinfo  {journal}
  {Quantum}\ }\textbf {\bibinfo {volume} {4}},\ \bibinfo {pages} {274}
  (\bibinfo {year} {2020})}\BibitemShut {NoStop}%
\bibitem [{\citenamefont {Salvia}\ and\ \citenamefont
  {Giovannetti}(2022)}]{salvia2022extracting}%
  \BibitemOpen
  \bibfield  {author} {\bibinfo {author} {\bibfnamefont {R.}~\bibnamefont
  {Salvia}}\ and\ \bibinfo {author} {\bibfnamefont {V.}~\bibnamefont
  {Giovannetti}},\ }\href@noop {} {\bibfield  {journal} {\bibinfo  {journal}
  {Physical Review A}\ }\textbf {\bibinfo {volume} {105}},\ \bibinfo {pages}
  {012414} (\bibinfo {year} {2022})}\BibitemShut {NoStop}%
\bibitem [{\citenamefont {\AA{}berg}(2014)}]{Aberg2014}%
  \BibitemOpen
  \bibfield  {author} {\bibinfo {author} {\bibfnamefont {J.}~\bibnamefont
  {\AA{}berg}},\ }\href {https://doi.org/10.1103/PhysRevLett.113.150402}
  {\bibfield  {journal} {\bibinfo  {journal} {Phys. Rev. Lett.}\ }\textbf
  {\bibinfo {volume} {113}},\ \bibinfo {pages} {150402} (\bibinfo {year}
  {2014})}\BibitemShut {NoStop}%
\bibitem [{\citenamefont {Vaccaro}\ \emph {et~al.}(2018)\citenamefont
  {Vaccaro}, \citenamefont {Croke},\ and\ \citenamefont
  {Barnett}}]{Vaccaro2018}%
  \BibitemOpen
  \bibfield  {author} {\bibinfo {author} {\bibfnamefont {J.~A.}\ \bibnamefont
  {Vaccaro}}, \bibinfo {author} {\bibfnamefont {S.}~\bibnamefont {Croke}},\
  and\ \bibinfo {author} {\bibfnamefont {S.~M.}\ \bibnamefont {Barnett}},\
  }\href {https://doi.org/10.1088/1751-8121/aac112} {\bibfield  {journal}
  {\bibinfo  {journal} {Journal of Physics A: Mathematical and Theoretical}\
  }\textbf {\bibinfo {volume} {51}},\ \bibinfo {pages} {414008} (\bibinfo
  {year} {2018})}\BibitemShut {NoStop}%
\bibitem [{\citenamefont {Lipka-Bartosik}\ and\ \citenamefont
  {Skrzypczyk}(2021{\natexlab{a}})}]{Lipka-Bartosik2021}%
  \BibitemOpen
  \bibfield  {author} {\bibinfo {author} {\bibfnamefont {P.}~\bibnamefont
  {Lipka-Bartosik}}\ and\ \bibinfo {author} {\bibfnamefont {P.}~\bibnamefont
  {Skrzypczyk}},\ }\href {https://doi.org/10.1103/PhysRevX.11.011061}
  {\bibfield  {journal} {\bibinfo  {journal} {Phys. Rev. X}\ }\textbf {\bibinfo
  {volume} {11}},\ \bibinfo {pages} {011061} (\bibinfo {year}
  {2021}{\natexlab{a}})}\BibitemShut {NoStop}%
\bibitem [{\citenamefont {Gallego}\ \emph {et~al.}(2016)\citenamefont
  {Gallego}, \citenamefont {Eisert},\ and\ \citenamefont
  {Wilming}}]{Gallego2016}%
  \BibitemOpen
  \bibfield  {author} {\bibinfo {author} {\bibfnamefont {R.}~\bibnamefont
  {Gallego}}, \bibinfo {author} {\bibfnamefont {J.}~\bibnamefont {Eisert}},\
  and\ \bibinfo {author} {\bibfnamefont {H.}~\bibnamefont {Wilming}},\ }\href
  {https://doi.org/10.1088/1367-2630/18/10/103017} {\bibfield  {journal}
  {\bibinfo  {journal} {New Journal of Physics}\ }\textbf {\bibinfo {volume}
  {18}},\ \bibinfo {pages} {103017} (\bibinfo {year} {2016})}\BibitemShut
  {NoStop}%
\bibitem [{\citenamefont {Boes}\ \emph
  {et~al.}(2019{\natexlab{a}})\citenamefont {Boes}, \citenamefont {Gallego},
  \citenamefont {Ng}, \citenamefont {Eisert},\ and\ \citenamefont
  {Wilming}}]{boes2019bypassing}%
  \BibitemOpen
  \bibfield  {author} {\bibinfo {author} {\bibfnamefont {P.}~\bibnamefont
  {Boes}}, \bibinfo {author} {\bibfnamefont {R.}~\bibnamefont {Gallego}},
  \bibinfo {author} {\bibfnamefont {N.~H.~Y.}\ \bibnamefont {Ng}}, \bibinfo
  {author} {\bibfnamefont {J.}~\bibnamefont {Eisert}},\ and\ \bibinfo {author}
  {\bibfnamefont {H.}~\bibnamefont {Wilming}},\ }\href@noop {} {\bibinfo
  {title} {By-passing fluctuation theorems}} (\bibinfo {year}
  {2019}{\natexlab{a}}),\ \Eprint {https://arxiv.org/abs/1904.01314}
  {arXiv:1904.01314} \BibitemShut {NoStop}%
\bibitem [{\citenamefont {Boes}\ \emph
  {et~al.}(2019{\natexlab{b}})\citenamefont {Boes}, \citenamefont {Eisert},
  \citenamefont {Gallego}, \citenamefont {Müller},\ and\ \citenamefont
  {Wilming}}]{Boes2019}%
  \BibitemOpen
  \bibfield  {author} {\bibinfo {author} {\bibfnamefont {P.}~\bibnamefont
  {Boes}}, \bibinfo {author} {\bibfnamefont {J.}~\bibnamefont {Eisert}},
  \bibinfo {author} {\bibfnamefont {R.}~\bibnamefont {Gallego}}, \bibinfo
  {author} {\bibfnamefont {M.~P.}\ \bibnamefont {Müller}},\ and\ \bibinfo
  {author} {\bibfnamefont {H.}~\bibnamefont {Wilming}},\ }\bibfield  {journal}
  {\bibinfo  {journal} {Physical Review Letters}\ }\textbf {\bibinfo {volume}
  {122}},\ \href {https://doi.org/10.1103/physrevlett.122.210402}
  {10.1103/physrevlett.122.210402} (\bibinfo {year}
  {2019}{\natexlab{b}})\BibitemShut {NoStop}%
\bibitem [{\citenamefont {Henao}\ and\ \citenamefont
  {Uzdin}(2022)}]{Henao2022}%
  \BibitemOpen
  \bibfield  {author} {\bibinfo {author} {\bibfnamefont {I.}~\bibnamefont
  {Henao}}\ and\ \bibinfo {author} {\bibfnamefont {R.}~\bibnamefont {Uzdin}},\
  }\href {https://doi.org/10.48550/ARXIV.2202.07192} {\bibinfo {title}
  {Catalytic leverage of correlations and mitigation of dissipation in
  information erasure}} (\bibinfo {year} {2022})\BibitemShut {NoStop}%
\bibitem [{\citenamefont {Henao}\ and\ \citenamefont
  {Uzdin}(2021)}]{Henao_2021}%
  \BibitemOpen
  \bibfield  {author} {\bibinfo {author} {\bibfnamefont {I.}~\bibnamefont
  {Henao}}\ and\ \bibinfo {author} {\bibfnamefont {R.}~\bibnamefont {Uzdin}},\
  }\href {https://doi.org/10.22331/q-2021-09-21-547} {\bibfield  {journal}
  {\bibinfo  {journal} {Quantum}\ }\textbf {\bibinfo {volume} {5}},\ \bibinfo
  {pages} {547} (\bibinfo {year} {2021})}\BibitemShut {NoStop}%
\bibitem [{\citenamefont {Lostaglio}\ and\ \citenamefont
  {M{\"u}ller}(2019)}]{lostaglio2019coherence}%
  \BibitemOpen
  \bibfield  {author} {\bibinfo {author} {\bibfnamefont {M.}~\bibnamefont
  {Lostaglio}}\ and\ \bibinfo {author} {\bibfnamefont {M.~P.}\ \bibnamefont
  {M{\"u}ller}},\ }\href@noop {} {\bibfield  {journal} {\bibinfo  {journal}
  {Physical review letters}\ }\textbf {\bibinfo {volume} {123}},\ \bibinfo
  {pages} {020403} (\bibinfo {year} {2019})}\BibitemShut {NoStop}%
\bibitem [{\citenamefont {Lipka-Bartosik}\ and\ \citenamefont
  {Skrzypczyk}(2021{\natexlab{b}})}]{PhysRevLett.127.080502}%
  \BibitemOpen
  \bibfield  {author} {\bibinfo {author} {\bibfnamefont {P.}~\bibnamefont
  {Lipka-Bartosik}}\ and\ \bibinfo {author} {\bibfnamefont {P.}~\bibnamefont
  {Skrzypczyk}},\ }\href {https://doi.org/10.1103/PhysRevLett.127.080502}
  {\bibfield  {journal} {\bibinfo  {journal} {Phys. Rev. Lett.}\ }\textbf
  {\bibinfo {volume} {127}},\ \bibinfo {pages} {080502} (\bibinfo {year}
  {2021}{\natexlab{b}})}\BibitemShut {NoStop}%
\bibitem [{\citenamefont {Jaynes}\ and\ \citenamefont
  {Cummings}(1963)}]{Jaynes1963}%
  \BibitemOpen
  \bibfield  {author} {\bibinfo {author} {\bibfnamefont {E.}~\bibnamefont
  {Jaynes}}\ and\ \bibinfo {author} {\bibfnamefont {F.}~\bibnamefont
  {Cummings}},\ }\href {https://doi.org/10.1109/PROC.1963.1664} {\bibfield
  {journal} {\bibinfo  {journal} {Proceedings of the IEEE}\ }\textbf {\bibinfo
  {volume} {51}},\ \bibinfo {pages} {89} (\bibinfo {year} {1963})}\BibitemShut
  {NoStop}%
\bibitem [{Note2()}]{Note2}%
  \BibitemOpen
  \bibinfo {note} {Taking more Fock states ($N$) does not change the
  qualitative aspects of our analysis. The only difference is that the number
  of different virtual temperatures will be generally proportional to $N$,
  leading to a more convoluted evolution picture.}\BibitemShut {Stop}%
\bibitem [{\citenamefont {Kinoshita}\ \emph {et~al.}(2006)\citenamefont
  {Kinoshita}, \citenamefont {Wenger},\ and\ \citenamefont
  {Weiss}}]{kinoshita2006quantum}%
  \BibitemOpen
  \bibfield  {author} {\bibinfo {author} {\bibfnamefont {T.}~\bibnamefont
  {Kinoshita}}, \bibinfo {author} {\bibfnamefont {T.}~\bibnamefont {Wenger}},\
  and\ \bibinfo {author} {\bibfnamefont {D.~S.}\ \bibnamefont {Weiss}},\
  }\href@noop {} {\bibfield  {journal} {\bibinfo  {journal} {Nature}\ }\textbf
  {\bibinfo {volume} {440}},\ \bibinfo {pages} {900} (\bibinfo {year}
  {2006})}\BibitemShut {NoStop}%
\bibitem [{\citenamefont {Brenes}\ \emph {et~al.}(2020)\citenamefont {Brenes},
  \citenamefont {LeBlond}, \citenamefont {Goold},\ and\ \citenamefont
  {Rigol}}]{Brenes2020}%
  \BibitemOpen
  \bibfield  {author} {\bibinfo {author} {\bibfnamefont {M.}~\bibnamefont
  {Brenes}}, \bibinfo {author} {\bibfnamefont {T.}~\bibnamefont {LeBlond}},
  \bibinfo {author} {\bibfnamefont {J.}~\bibnamefont {Goold}},\ and\ \bibinfo
  {author} {\bibfnamefont {M.}~\bibnamefont {Rigol}},\ }\href
  {https://doi.org/10.1103/PhysRevLett.125.070605} {\bibfield  {journal}
  {\bibinfo  {journal} {Phys. Rev. Lett.}\ }\textbf {\bibinfo {volume} {125}},\
  \bibinfo {pages} {070605} (\bibinfo {year} {2020})}\BibitemShut {NoStop}%
\bibitem [{\citenamefont {Anderson}(1958)}]{anderson1958absence}%
  \BibitemOpen
  \bibfield  {author} {\bibinfo {author} {\bibfnamefont {P.~W.}\ \bibnamefont
  {Anderson}},\ }\href@noop {} {\bibfield  {journal} {\bibinfo  {journal}
  {Physical review}\ }\textbf {\bibinfo {volume} {109}},\ \bibinfo {pages}
  {1492} (\bibinfo {year} {1958})}\BibitemShut {NoStop}%
\bibitem [{\citenamefont {Abanin}\ \emph {et~al.}(2019)\citenamefont {Abanin},
  \citenamefont {Altman}, \citenamefont {Bloch},\ and\ \citenamefont
  {Serbyn}}]{Abanin_2019}%
  \BibitemOpen
  \bibfield  {author} {\bibinfo {author} {\bibfnamefont {D.~A.}\ \bibnamefont
  {Abanin}}, \bibinfo {author} {\bibfnamefont {E.}~\bibnamefont {Altman}},
  \bibinfo {author} {\bibfnamefont {I.}~\bibnamefont {Bloch}},\ and\ \bibinfo
  {author} {\bibfnamefont {M.}~\bibnamefont {Serbyn}},\ }\bibfield  {journal}
  {\bibinfo  {journal} {Reviews of Modern Physics}\ }\textbf {\bibinfo {volume}
  {91}},\ \href {https://doi.org/10.1103/revmodphys.91.021001}
  {10.1103/revmodphys.91.021001} (\bibinfo {year} {2019})\BibitemShut {NoStop}%
\bibitem [{\citenamefont {Nandkishore}\ and\ \citenamefont
  {Huse}(2015)}]{nandkishore2015many}%
  \BibitemOpen
  \bibfield  {author} {\bibinfo {author} {\bibfnamefont {R.}~\bibnamefont
  {Nandkishore}}\ and\ \bibinfo {author} {\bibfnamefont {D.~A.}\ \bibnamefont
  {Huse}},\ }\href@noop {} {\bibfield  {journal} {\bibinfo  {journal} {Annu.
  Rev. Condens. Matter Phys.}\ }\textbf {\bibinfo {volume} {6}},\ \bibinfo
  {pages} {15} (\bibinfo {year} {2015})}\BibitemShut {NoStop}%
\bibitem [{\citenamefont {Lostaglio}(2016)}]{Lostaglio2016}%
  \BibitemOpen
  \bibfield  {author} {\bibinfo {author} {\bibfnamefont {M.}~\bibnamefont
  {Lostaglio}},\ }\href@noop {} {\emph {\bibinfo {title} {The resource theory
  of quantum thermodynamics}}}\ (\bibinfo  {publisher} {Imperial College
  London},\ \bibinfo {year} {2016})\BibitemShut {NoStop}%
\bibitem [{\citenamefont {\ifmmode \acute{C}\else
  \'{C}\fi{}wikli\ifmmode~\acute{n}\else \'{n}\fi{}ski}\ \emph
  {et~al.}(2015)\citenamefont {\ifmmode \acute{C}\else
  \'{C}\fi{}wikli\ifmmode~\acute{n}\else \'{n}\fi{}ski}, \citenamefont
  {Studzi\ifmmode~\acute{n}\else \'{n}\fi{}ski}, \citenamefont {Horodecki},\
  and\ \citenamefont {Oppenheim}}]{Cwiklinski2015}%
  \BibitemOpen
  \bibfield  {author} {\bibinfo {author} {\bibfnamefont {P.}~\bibnamefont
  {\ifmmode \acute{C}\else \'{C}\fi{}wikli\ifmmode~\acute{n}\else
  \'{n}\fi{}ski}}, \bibinfo {author} {\bibfnamefont {M.}~\bibnamefont
  {Studzi\ifmmode~\acute{n}\else \'{n}\fi{}ski}}, \bibinfo {author}
  {\bibfnamefont {M.}~\bibnamefont {Horodecki}},\ and\ \bibinfo {author}
  {\bibfnamefont {J.}~\bibnamefont {Oppenheim}},\ }\href
  {https://doi.org/10.1103/PhysRevLett.115.210403} {\bibfield  {journal}
  {\bibinfo  {journal} {Phys. Rev. Lett.}\ }\textbf {\bibinfo {volume} {115}},\
  \bibinfo {pages} {210403} (\bibinfo {year} {2015})}\BibitemShut {NoStop}%
\end{thebibliography}%

\appendix
\onecolumngrid
\section{Effective temperatures of single-copy quantum systems}
\label{app:1}
We focus on the effective temperature $T_c(A)$ as the analysis for the complementary case $T_h(A)$ is similar. Conservation of energy in the form of $[U, H_A + H_B] = 0$ implies $Q(T, H_B, U) = \tr[H_A (\rho_S-\mathcal{E}_{T}(\rho_S))]$, and so Eq.  (5) can be written as
\begin{equation}
\begin{aligned}
\label{eq:5}
T_c(A) :=\, \min_{\mathcal{E}_{T}} \quad & T  \\
\textrm{s.t.} \quad & \Tr[H_A(\mathcal{E}_{T}(\rho_A) - \rho_A)] > 0,\\
\quad & [U, H_A+H_B] = 0.
\end{aligned}
\end{equation}
In the above we used the fact that optimization over Hamiltonians $H_B$ and energy-preserving unitaries $U$ is by definition equivalent to optimizing over all quantum channels of the form (6). This is a well-studied class of quantum channels known as thermal operations \cite{Horodecki2013,Brand_o_2015,Lostaglio2019}. We will often use this correspondence to simplify the expressions for effective temperatures.

The commutation relation $[U, H_A+H_B] = 0$ imposes strong limitations on the evolution of quantum states. These restrictions can be conveniently understood using the notion of Bohr spectrum. That is, for a given Hamiltonian $H$ with spectrum $\bm{\lambda}(H) = \{{\epsilon_i}\}$ we define the \emph{Bohr spectrum} $\Omega(H)$ to be the set of transition frequencies $\omega \in \Omega(H)$ such that $\omega = \epsilon_i - \epsilon_j$ for some $\epsilon_i, \epsilon_j \in \bm{\lambda}(H)$. Consequently, any density operator $\rho$ can be written as
\begin{align}
    \rho = \sum_{\omega \in \Omega(H)} \rho(\omega),
\end{align}
where $\rho(\omega)$ are called \emph{modes of coherence} \cite{lostaglio2015description,Lostaglio2016}. It can be verified that $[U, H_A + H_B] = 0$ implies that the effective channels of the form (6) are symmetric under time translations, that is $e^{-i H t} \mathcal{E}_{\beta}(\rho) e^{i H t} =  \mathcal{E}_{\beta}(e^{-i H t} \rho e^{i H t})$ holds for all $t$. An important property of such channels is that each mode of coherence evolves independently, which
allows to formulate the constraints imposed by unitarity and energy-conservation on quantum evolution in terms of independent transformation laws for each mode \cite{lostaglio2015description}. In other words, $\mathcal{E}_{T}(\rho) = \sigma$ for some channel $\mathcal{E}_T$ if and only if 
\begin{align}
    \mathcal{E}_{T} (\rho(\omega)) = \sigma(\omega) \qquad \text{for all}\quad \omega \in \Omega(H).
\end{align}
Although it is notoriously difficult to characterize how different modes $\rho(\omega)$ evolve under $\mathcal{E}_{T}$ \cite{Cwiklinski2015,Lostaglio2019}, the conditions for transforming the zeroth mode $\rho(0)$, i.e. the diagonal of $\rho$ in the basis of $H$, have been fully characterized \cite{Horodecki2013}. This will be enough for our purposes, and we will express these conditions using Gibbs-stochastic matrices, i.e stochastic matrices $G_{T}$ that satisfy $G_{T} \bm{g} = \bm{g}$ for $\bm{g} = \text{diag}[\gamma(T, H_A)]$ being a vector of Gibbs weights \cite{Janzing2000}. Such matrices characterize energy flows induced by channels $\mathcal{E}_{T}$ and provide partial information about the effective dynamics on the system.

Denoting $\bm{p} = \text{diag}[\rho_A(0)]$ we can observe that $\Tr[H_A \mathcal{E}_{T}(\rho_A)] = \Tr\big[H_A \mathcal{E}_{T}(\rho_A{(0)})\big]$ and consequently
\begin{align}
    \Tr[H_A(\mathcal{E}_{T}(\rho_A) \!- \!\rho_A)] = \bm{\lambda}(H_A)^T (G_{T} - 1) \mathbf{p},
\end{align}
where $\bm{\lambda}(H)$ should be viewed as a vector of eigenvalues of $H$. The effective temperature $T_c(A)$ can be then written as
\begin{equation}
\begin{aligned}
\label{eq:app_th}
T_c(A) := \quad \min_{G_{T}} \quad & T \\
\textrm{s.t.} \quad & \bm{\lambda}(H_A)^T (G_{T}-1) \mathbf{p} > 0, \\
\quad & G_{T} \quad \text{is Gibbs-stochastic}.
\end{aligned}
\end{equation}
Similarily we can write the expression for the effective temperature $T_h(A)$ as
\begin{equation}
\begin{aligned}
T_h(A) := \quad \max_{G_{T}} \quad & T \\
\textrm{s.t.} \quad & \bm{\lambda}(H_A)^T (G_{T}-1) \mathbf{p} < 0, \\
\quad & G_{T} \quad \text{is Gibbs-stochastic}.
\end{aligned}
\end{equation}

\noindent For any quantum system $A$ we define the \emph{virtual temperature spectrum} (VTS) to be the set 

\begin{align}
    \mathbb{V}(A) = \{T_{ij}\, |\, 1 \leq i, j \leq d_A \,\, \text{and} \,\, \langle i |H_A |i \rangle \neq \langle j |H_A |j \rangle \},
\end{align}
where $T_{ij} := ({\epsilon_j - \epsilon_i}) / \log(p_i/p_j)$ with $\epsilon_i := \langle i |H_A |i \rangle$. 

Let us now denote the minimal element of $\mathbb{V}(A)$ with $T_{\text{min}}$. First we will show that $T_c(A) \geq T_{\text{min}}$, i.e. there is no system $B$ that can be cooled down using system $A$ when $T$ is smaller than the minimal virtual temperature of $A$. Then we will construct an explicit protocol that allows a qubit thermometer to be cooled as long as $T$ is larger than the minimal virtual temperature of $A$, showing that $T_c(A) \leq T_{\text{min}}$. This will prove that $T_{\text{min}} = T_c(A)$ and solve the problem from Eq. (\ref{eq:app_th}).

We begin by proving a lower bound on $T_c(A)$, that is $T_c(A) \geq T_{\text{min}}$.  By contradiction, let us assume that the temperature of the thermal probe, $T$, satisfies $T < T_{\text{min}}$. This implies $\bm{\lambda}(H)^T \bm{p} \geq \bm{\lambda}(H)^T G \bm{p}$ for all Gibbs-stochastic matrices $G$. To see this, let us introduce a matrix $M$ with elements $M_{ij}$ such that $G_{ij} = M_{ij} e^{-\beta (\epsilon_j - \epsilon_i)}$. Consider

\begin{align}
    \bm{\lambda}(H)^T G \bm{p} = \sum_{i, j} p_i G_{ij} \epsilon_j =  \sum_{i,j} p_i M_{ij} e^{-\beta(\epsilon_j - \epsilon_i)} \epsilon_j \leq \sum_{i,j} p_i M_{ij} e^{-\beta_{ij}(\epsilon_j - \epsilon_i)} \epsilon_j = \sum_{i,j} p_j M_{ij} \epsilon_j,
\end{align}
\chg{where the quantities $\beta_{ij} = 1 / T_{ij}$ are the inverse virtual temperatures.} By construction matrix $M$ preserves the identity vector, i.e. $\sum_{i} M_{ij} = 1$, which follows from the Gibbs-stochastic property of $G$. This allows us to write
\begin{align}
    \sum_{i,j} p_j M_{ij} \epsilon_j = \sum_{j} p_j \epsilon_j = \bm{\lambda}(H)^T p.
\end{align}
In other words, when all virtual temperatures (and specifically $T < T_{\text{min}}$) are higher than $T$, the system is in a state with the maximal energy possible under all energy-preserving interactions with any thermal state at temperature $T$. Consequently, it is impossible to further increase its energy, or cool any thermometer. Therefore, the only possibility to cool the thermometer can occur when  $T_c(A) \geq T_{\text{min}}$, as claimed.

To prove the lower bound, suppose that $T_{\text{min}}$ corresponds to some energy levels $i$ and $j$, that is $T_{\text{min}} = (\epsilon_j - \epsilon_i) /\log(p_i/p_j)$. Consider a thermometer $B = (\mathcal{H}_B, H_B, \gamma_B)$ to be a qubit with $H_B = \text{diag}(0, \Delta)$ and $\gamma_B = \text{diag}(g_0, g_1)$, where $\Delta = \epsilon_j - \epsilon_i$ and $g_0 = (1+e^{-\beta \Delta})^{-1} = 1 - g_1$. The unitary $U$ is chosen to be a swap between degenerate energy levels, i.e.
\begin{align}
    \ket{\epsilon_i}_{A} \ot \ket{1}_B \xleftrightarrow{\quad U \quad} \ket{\epsilon_j}_A \ot \ket{0}_B. 
\end{align}
This unitary can be generated by a time-independent interaction Hamiltonian
\begin{align}
    H_{\text{int}} = \dyad{\epsilon_j}{\epsilon_i}_A \ot \dyad{0}{1}_B + \dyad{\epsilon_i}{\epsilon_j}_A \ot \dyad{1}{0}_B 
\end{align}
acting for time $t = \pi/2$, i.e. $U = e^{-i H_{\text{int}}t}$. Furthermore,  this particular interaction satisfies $[H_{\text{int}}, H_A \ot 1_B + 1_A + H_B] = 0$ as desired. It can be verified that $U$ changes thermometer's occupations as $(g_0, g_1) \rightarrow (g_0 + \delta_{ij}, g_1-\delta_{ij})$ with $\delta_{ij} := p_i g_0e^{-\beta \Delta}(1 - e^{-(\beta_{\text{max}}-\beta)\Delta})$. 
Therefore, as long as $T \geq T_{\text{min}}$ we have $\delta_{ij} > 0$ and cooling can be achieved, leading to $T_c(A) \leq T_{\text{min}}$ as claimed.

We found the explicit solution to the optimization problem (\ref{eq:app_th}). The proof for the effective hot temperature $T_h(A)$ proceeds identically, so we omit it here. To summarize, for a quantum system $A = (H_A, \rho_A)$ with a virtual temperature spectrum $\mathbb{V}(A) = \{T_{ij}\}_{i\neq j}$, the effective temperatures are given by

\begin{align}
    \label{app:tc}
    T_c(A) &=  \min \mathbb{V}(A) = \max_{(i,j): \,\, i \neq j} T_{ij}, \\
    \label{app:th}
    T_h(A) &= \max \mathbb{V}(A) = \min_{(i,j): \,\, i \neq j} T_{ij}.
\end{align}

\section{Effective temperatures of macroscopic quantum systems}
\label{app:2}
In this Appendix we solve the optimization problem defining effective temperatures of quantum systems processed collectively. We then prove the form of expansions with respect to parameter $\delta$. Since the reasoning for both $T_c^{\infty}(A,\delta)$ and $T_h^{\infty}(A,\delta)$ is completely analogous, we focus on the former. 

\subsection{Solution of the problem defining $T_c^{\infty}(A,\delta)$.}
The effective temperature $T_c^{\infty}(A,\delta)$ is given by
\begin{equation}
\begin{aligned}
    \label{eq:t_inf_opt_prob}
T_c^{\infty}(A, \delta) := \lim_{n \rightarrow \infty} T_c(A^n,n\delta) =\, \lim_{n \rightarrow \infty} \min_{\mathcal{E}_{T}} \quad & \,\,T  \\
\textrm{s.t.} \quad & \Tr[H_A^{\ot n} (\mathcal{E}_{T}(\rho_A^{\ot n}) - \rho_A^{\ot n})] \geq n \delta,\\
\quad & [U, H_A^{\ot n}+H_B] = 0,
\end{aligned}
\end{equation}
We want to show that the optimal solution of the above problem is exactly the same as the solution to the much simpler problem 
\begin{align}
\label{eq:equiv_problem}
\min_{\sigma_A} \quad & T  \\ \label{eq:con_1}
\textrm{s.t.} \quad & \Tr[H_A(\sigma_A - \rho_A)] \geq \delta,\\
\label{eq:con_2}
\quad & F_T(\rho_A, H_A) \geq F_T(\sigma_A, H_A),
\end{align}
where $F(\rho, H) := \Tr[\rho H] - T S(\rho)$ is the non-equilibrium free energy. Then we will show that the equivalent problem (\ref{eq:equiv_problem}) can be solved explicitly.

We begin by showing that for any $n \in \mathbb{N}$ the constraints 
\begin{align}
    \Tr[U, H_A^{\ot n}+ H_B] &= 0, \\
    \Tr[H_A^{\ot n}\left(\mathcal{E}_T(\rho_A^{\ot n}) - \rho_A^{\ot n}\right)] &\geq n\delta,
\end{align}
imply $F_T(\rho_A, H_A) \geq F_T(\sigma_A, H_A)$ for any $H_B$ and $U$. To see this, notice that the unitary invariance of von Neuman entropy implies
\begin{align}
    \Delta S_S + \Delta S_B \geq 0.
\end{align}
Since $\Delta S_B = -\beta \Delta E_S - D(\rho_B'\| \gamma_B) \leq -\beta \Delta E_S$, we can write 
\begin{align}
     F_T(\rho_A, H_A) \geq F_T(\sigma_A, H_A).
\end{align}
Let us now assume
\begin{align}
    \label{eq:ac_second_dirA}
    F_T(\rho_A, H_A) &\geq F_T(\sigma_A, H_A), \\
    \label{eq:ac_second_dirB}
    \Tr[H_A(\sigma_A-\rho_A)] &\geq \delta
\end{align}
We shall prove that for sufficiently large $n \in \mathbb{N}$ there always exist a choice of operators $H_B$ and $U$ that achieves the optimum $T_c(A^n, n\delta)$.

Before we proceed we adapt the key result from Ref. \cite{Brand_o_2015} that provides a partial asymptotic characterisation of channels $\mathcal{E}_T$.
\begin{lemma}[Brandao et. al. \cite{Brand_o_2015}]
\label{lem:1}
Let $\rho_A$ be an arbitrary density matrix and $\sigma_A$ be an arbitrary block-diagonal density matrix. Then, for any $\epsilon > 0$ there is an $n\in \mathbb{N}$, a system $B = (H_B, \gamma_B(H_B, T))$ and a unitary $U$ satisfying $[U, H_A^{\ot n} + H_B] = 0$ and implementing a quantum channel $\mathcal{E}_T(\cdot) := \Tr_B\left[U(\cdot \ot \gamma_B(H_B, T))U^{\dagger}\right]$ such that
\begin{align}
    \label{eq:norm_eq}
    \norm{\mathcal{E}_T(\rho_A^{\ot n}) - \sigma_A^{\ot n}}_1 \leq \epsilon
\end{align}
if and only if $F_T(\rho_A, H) \geq F_T(\sigma_A, H)$. Moreover, $\epsilon$ can be taken to be $\sim \mathcal{O}(e^{-a n})$ for some $a \in \mathbb{R}$. 
\end{lemma}
In other words, our assumption (\ref{eq:ac_second_dirA}) implies that for a sufficiently large $n$ there always exist a thermometer system $B=(H_B, \gamma(H_B,T))$ and an energy-conserving unitary $U$ that implements a channel $\mathcal{E}_T$ such that $\mathcal{E}_T(\rho_A^{\ot n})$ is arbitrarily close to $\sigma^{\ot n}_A$ and $F_T(\rho_A, H_A) \geq F_T(\sigma_A, H_A)$. This implies that the second line of constraints in Eq. (\ref{eq:t_inf_opt_prob}) is satisfied. Moreover, the second assumption (\ref{eq:ac_second_dirB}) implies that for any $n \in \mathbb{N}$
\begin{align}
    \Tr[H_A^{\ot n}\left(\sigma_A^{\ot n} - \rho_A^{\ot n}\right)] \geq n \delta.
\end{align}
Moreover, Eq. (\ref{eq:norm_eq}) implies that $\Tr\left[M (\mathcal{E}_T(\rho_A^{\ot n}) - \sigma_A^{\ot n})\right] \leq \epsilon$ for any operator operator satisfying $0 \leq M \leq \mathbb{1}$. This, on the other hand, implies 
\begin{align}
    \mathcal{E}_T(\rho_A^{\ot n})  \leq \sigma_A^{\ot n} + \epsilon \mathbb{1}_A,
\end{align}
Therefore, for a sufficiently large $n$, the second term vanishes and we recover the first line of constraints in Eq. (\ref{eq:t_inf_opt_prob}). This means that the two problems, i.e. (\ref{eq:t_inf_opt_prob}) and (\ref{eq:equiv_problem}--\ref{eq:con_2}) have the same solutions.

Now we will show that the solution of the problem (\ref{eq:equiv_problem}) is given by 
\begin{align}
    \label{eq:T_opt_asymp}
    T_{\text{opt}} = \frac{\delta}{S(\gamma_A(E+\delta)) - S(\rho_A)},
\end{align}
where $\gamma_A(E+\delta) := e^{-\beta^*(E+\delta) H_A} / Z^*(E+\delta)$ with $Z^*(E+\delta) := \Tr e^{-\beta^*(E+\delta) H_A}$ and $\beta^*(E+\delta)$ is chosen such that the following implicit relation is satisfied
\begin{align}
    \Tr[H_A \rho_A] + \delta = \Tr[H_A \gamma_A(E+\delta)].
\end{align}
To prove that Eq. (\ref{eq:T_opt_asymp}) describes the solution to our simplified problem (\ref{eq:equiv_problem}) we choose a potentially sub-optimal guess for the state $\sigma_A$, i.e. $\sigma_A = \gamma_A(E+\delta)$. It is easy to verify that this choice is feasible and achieves the value from Eq. (\ref{eq:T_opt_asymp}). To see that this is the minimal value achievable under the constraints, notice that any feasible state $\sigma_A$ must satisfy both
\begin{align}
    T \geq \frac{\delta}{S(\sigma_A) - S(\rho_A)} \qquad \text{and} \qquad  \Tr[H_A\sigma_A] \geq \Tr[H_A \rho_A] + \delta.
\end{align}
Notice that the optimal $T$ is always positive. In order to maximize $S(\sigma_A)$ we further note that Gibbs states have maximal entropy for a given energy. Since for Gibbs states with non-negative temperature entropy decreases with average energy, it is best to choose $\Tr[H_A\sigma_A] = \Tr[H_A\rho_A]+\delta$.  Therefore, we conclude that Eq. (\ref{eq:T_opt_asymp}) is the solution to our initial problem. 

\subsection{The form of $T_c^{\infty}(A,\delta)$ and $T_h^{\infty}(A,\delta)$ for small $\delta$.}

Let us now expand the effective temperatures for small values of $\delta$. We start by expanding the von Neumann entropy
\begin{align}
    S(\gamma_A(E+\delta)) :=  S_{\gamma_A}(E+\delta) =  S_{\gamma_A}(E) + \frac{\partial S_{\gamma_A}(E)}{\partial E} \delta + \frac{1}{2} \frac{\partial^2 S_{\gamma_A}(E)}{\partial^2 E} \delta^2 + \mathcal{O}(\delta^3).
\end{align}
Plugging this expression into the formula for $T_c^{\infty}(A,\delta)$ yields
\begin{align}
    \label{eq:beta_c_expansion}
    \beta_{c}(A, \delta) = \frac{S_{\gamma_A}(E) - S(\rho_A)}{\delta} + \frac{\partial S_{\gamma_A}(E)}{\partial E} + \frac{1}{2} \frac{\partial^2 S_{\gamma_A}(E)}{\partial^2 E} \delta + \mathcal{O}(\delta^2).
\end{align}
In order to find the derivatives of $S_{\gamma_A}(E)$ with respect to $E$ notice that we have
\begin{align}
    \frac{\partial \log Z^*(E)}{\partial E} = \frac{\partial \log Z^*(E)}{\partial \beta^*(E)} \frac{\partial \beta^*(E)}{\partial E} = -E  \frac{\partial \beta^*(E)}{\partial E}.
\end{align}
Moreover, differentiating both sides of $\Tr[H_A\gamma^*(\beta^*, E)] = E$ with respect to $E$ we find
\begin{align}
    \frac{\partial \beta^*(E)}{\partial E} =  - \frac{1}{\Delta^2 E(\gamma_A)}.
\end{align}
Using the above expressions we can write
\begin{align}
    \frac{\partial S_{\gamma_A}(E)}{\partial E} &= E  \frac{\partial \beta^*(E)}{\partial E} + \beta^*(E) +  \frac{\partial \log Z^*(E)}{\partial E} = \beta^*(E),\\
    \frac{\partial^2 S_{\gamma_A}(E)}{\partial^2 E} &= \frac{\partial \beta^*(E)}{\partial E} =  - \frac{1}{\Delta^2 E(\gamma_A)}.
\end{align}
Plugging in the above into our expression from Eq. (\ref{eq:beta_c_expansion}) gives the desired result.

\section{Case study: Effective temperatures of a qutrit system}
 \label{app:3}
\begin{figure*}[]
    \centering
    \includegraphics[width=\linewidth]{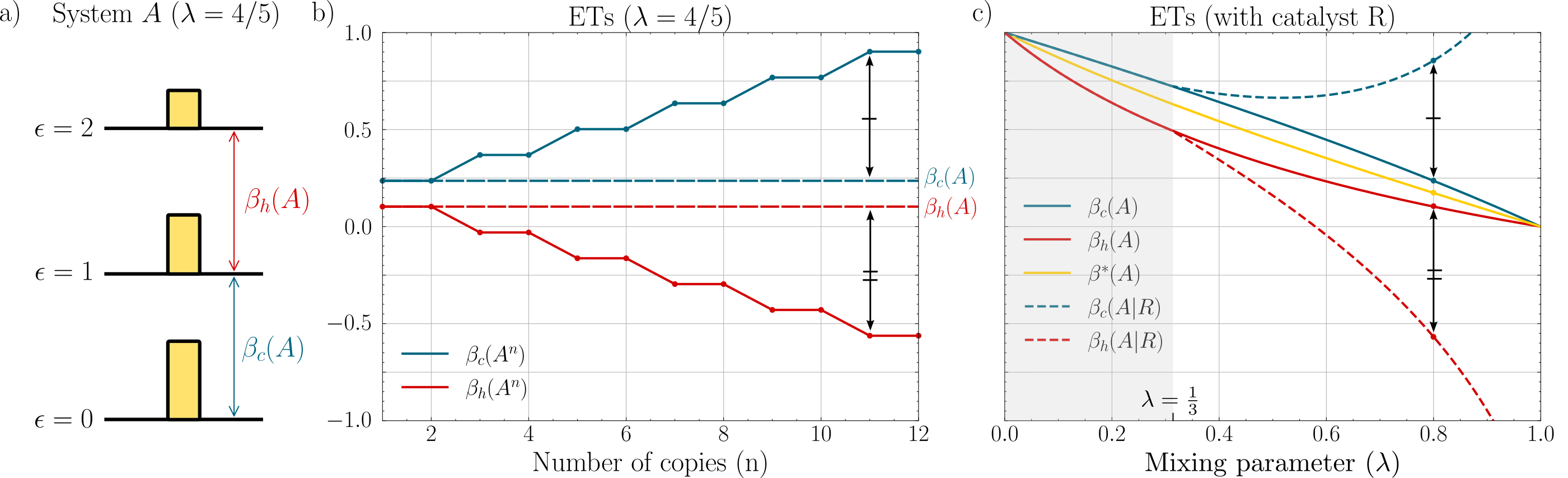}
    \caption{\chg{Illustrative example. Panel $(a)$ shows the qutrit $A$ and its effective \chg{inverse} temperatures (ETs) $\beta_{c/h}(A)$. Panel $(b)$ shows the effective \chg{inverse} temperatures as a function of the number of copies $n$ of $A$. As $n$ increases, the range of $\beta_{c/h}(A^n)$ broadens. Panel $(c)$ shows effective \chg{inverse} temperatures in the presence of a qubit catalyst (reference frame). Notice that the effective \chg{inverse} temperatures $\beta_{c/h}(A|R)$ depend on the coherence in the state $\rho_A$. When $\lambda \leq 1/3$, there is no catalytic advantage, while the case $\lambda = 4/5$ is equivalent to having $n = 11$ copies of $A$. }}
    \label{fig:2}
\end{figure*}

\noindent Consider a qutrit $A = (\rho_A, H_A)$, shown pictorially in Fig. \ref{fig:2}, with $H_A = \dyad{\epsilon_1} + 2\dyad{\epsilon_2}$ and
\begin{align}
    \rho_A = (1-\lambda) \gamma_A(H_A, \beta)+ \lambda \dyad{\psi},
\end{align}
where $\ket{\psi} = (\ket{\epsilon_0}+\ket{\epsilon_1}+\ket{\epsilon_2})/\sqrt{3}$ and $0 \leq \lambda \leq 1$.  In Fig. \ref{fig:2} we plot the effective temperatures $T_{c/h}(A)$ as computed from Eqs. (\ref{app:tc}) and (\ref{app:th}). In particular, $A$ can cool some thermometer $B$ at temperature $T$ as long as $T < T_c(A)$. The thermometer can be a qubit with $H_B = \dyad{\epsilon_1}_B$ coupled to energy levels $\ket{\epsilon_0}_A$ and $\ket{\epsilon_1}_A$ via Hamiltonian $H_{\text{int}}^{\text{c}} = \dyad{\epsilon_1}{\epsilon_0}_A \ot \dyad{\epsilon_0}{\epsilon_1}_B + \text{h.c.}$, acting for time $t = \pi/2$ so that $U = e^{-i H_{\text{int}}^{\text{c}}t}$. Similarly, $A$ can heat up $B$ as long as $T > T_h(A)$, which can be achieved by coupling $B$ to energy levels $\ket{\epsilon_1}_A$ and $\ket{\epsilon_2}_A$ via $H_{\text{int}}^{\text{h}} = \dyad{\epsilon_2}{\epsilon_1}_A \ot \dyad{\epsilon_0}{\epsilon_1}_B + \text{h.c.}$ 

When $n$ copies of $A$ are used, the range of \chg{(inverse)} effective temperatures increases with $n$ (see Fig. \ref{fig:2}b). Still, the heat delivered from the multi-copy system scales sublinearly with $n$, and for large $n$ becomes negligible. In order to consider physically-relevant processes, we then require that at least $n \delta$ of energy is transferred as heat, as shown in Eq. (\ref{eq:t_inf_opt_prob}).

In order to analyse the effect of coherence activation using a catalyst, let us first consider $\beta = 0$ in which case energy populations of $\rho_A$ correspond to a thermal state at infinite temperature, i.e. $\gamma_A(H_A, \infty) = \mathbb{1}/3$, and the effective temperatures coincide, $\beta_c(A) = \beta_h(A) = 0$. We further assume access to a qubit system (catalyst) $R = (\phi_R, H_R)$ with $H_R = \dyad{\epsilon_1}$ and
\begin{align}
\phi_R =\frac{1}{2} \left(1-\frac{1}{\sqrt{2}}\right) \mathbb{1}_2+ \frac{1}{\sqrt{2}} \dyad{+},
\end{align}
where $\ket{+} = (\ket{\epsilon_0}+\ket{\epsilon_1})/\sqrt{2}$. As we will see, this ancillary quantum system acts as a reference frame. More specifically, it allows for implementing an effective dynamics on system $A$ that is not energy-conserving in the strict sense on system $A$, but only on the total system $A$ and $R$. This is itself not a surprise, as one can observe a similar behaviour when implementing quantum gates \cite{Aberg2014} using energy-conserving interactions. The surprising thing is that by fine-tuning the interaction and the state of the reference frame, this procedure can be implemented at virtually no back-action on the reference frame. This last fact is a key observation that allows us to associate the temperature with system $A$ only. That is, the ancillary system $R$ can never be used as a source of non-equilibrium itself.

The joint system $AR$ has two degenerate energy eigenspaces spanned, respectively, by $\{\ket{\epsilon_0\epsilon_1}, \ket{\epsilon_1\epsilon_0}\}$ and $\{\ket{\epsilon_2\epsilon_0}, \ket{\epsilon_1\epsilon_1}\}$. The protocol we consider consists of a \emph{preprocessing} step in which system $A$ interacts resonantly with $R$ in these degenerate subspaces, and a \emph{cooling/heating} step where the actual transfer of heat between $A$ and $B$ is carried out. \chg{For preprocessing, we consider a unitary $V$ defined as
\begin{align}
    V = \frac{1}{\sqrt{2}} \left(\mathbb{1} + i \ket{\epsilon_0\epsilon_1}\bra{\epsilon_1 \epsilon_0} + i \ket{\epsilon_1\epsilon_1}\bra{\epsilon_2 \epsilon_0} - i \ket{\epsilon_2\epsilon_0}\bra{\epsilon_1 \epsilon_1} - i \ket{\epsilon_1\epsilon_0}\bra{\epsilon_0 \epsilon_1}\right)
\end{align}
In particular, in the subspaces of degenerate energy, that is subspaces spanned by $\{\ket{\epsilon_0\epsilon_1}, \ket{\epsilon_1\epsilon_0}\}$ and $\{\ket{\epsilon_{2}\epsilon_0}, \ket{\epsilon_1 \epsilon_1}\}$, the unitary $V$ acts as
\begin{align}
    \label{eq:unitary_example1}
    V \ket{\epsilon_0\epsilon_1} &= \frac{\ket{\epsilon_0\epsilon_1} - i \ket{\epsilon_1\epsilon_0}}{\sqrt{2}}, \qquad  V \ket{\epsilon_1\epsilon_0} = \frac{\ket{\epsilon_1\epsilon_0} + i \ket{\epsilon_0\epsilon_1}}{\sqrt{2}}, \\
    \label{eq:unitary_example2}
    V \ket{\epsilon_2\epsilon_0} &= \frac{\ket{\epsilon_2\epsilon_0} + i \ket{\epsilon_1\epsilon_1}}{\sqrt{2}}, \qquad V \ket{\epsilon_1\epsilon_1} = \frac{-i \ket{\epsilon_2\epsilon_0}+ \ket{\epsilon_1\epsilon_1}}{\sqrt{2}}.
\end{align}}
Clearly, $[V, H_{AR}] = 0$ and $V$ rotates the initial state into $\sigma_{AR} = V \rho_A \ot \phi_R V^{\dagger}$ that leads to the reduced state
\begin{align}
\sigma_{A} = 
    \label{app:eq_prep_A}
    \frac{1}{12} 
    \begin{pmatrix}
    4+\sqrt{2} & x_- & 2x_+ \\
    x_- & 4-2 \sqrt{2} & x_- \\
    2x_+ & x_- & 4 + \sqrt{2}
    \end{pmatrix},
\end{align}
with $x_{\pm} = (\sqrt{2}\pm 1)/2$. Since the marginal state of the catalyst remains the same, $\sigma_R = \phi_R$, the process can be viewed as a correlated-catalytic transformation \cite{Gallego2016}. Notice that the state of $A$ is now different from the Gibbs state at infinite temperature: We redistributed its energy population using its internal coherence. Using Eqs. (\ref{app:tc}) and (\ref{app:th}) we can find the effective temperatures of the preprocessed state from Eq. (\ref{app:eq_prep_A}). A direct computation
reveals that the effective temperatures of $\sigma_A$, denoted $\beta_{c/h}(A|R)$, are given by 
\begin{align}
    \beta_{c/h}(A|R) := \pm \log\left(\frac{5}{2} + \frac{3}{\sqrt{2}}\right) \approx \pm 1.53.
\end{align}
The qubit reference frame $R$ therefore enables manipulation of energy coherence stored in $A$ to boost its cooling and heating capabilities. Notice that none of this would be possible without either initial coherence of $A$ (as demonstrated by no-broadcasting theorems \cite{Marvian2019,lostaglio2019coherence}), or without a coherent quantum catalyst (since for a diagonal $\phi_R$ the effective channel on $A$ will be time-symmetric). Moreover, a very similar protocol can be used to demonstrate this effect for all values of $\beta$ (see Fig. 2). In this case the same preprocessing step is applied, and one only needs to tune the initial state of the reference frame so that  $\sigma_R = \phi_R$ still holds.

This is reminiscent of the seminal work of Aberg showing that an energy-conserving unitary acting on a qubit and a harmonic oscillator can simulate any unitary on the qubit \cite{Aberg2014}. Aberg's protocol, however, requires the energy of the harmonic oscillator to change \cite{Vaccaro2018}, meaning that both systems $A$ and $R$ would be responsible for heating up/cooling down $B$. In our case catalysis ensures that it is only the energy of the system $A$ that is responsible for generating the flow of heat. This indicates a genuinely quantum mechanism that exploits energy coherence to generate the desired flow of heat. 

\section{Catalytic effective temperatures}
\label{app:4}
In this Appendix we define catalytic effective temperatures from the main text and show that they can be expressed as stated in Eq. (14). Notice that to prove this, it is enough to show that the optimal solution of the original problem is the same as the solution to the simplified problem defined in Eqs. (\ref{eq:equiv_problem}--\ref{eq:con_2}). 

Consider the extended scenario with a catalyst, i.e. consider a unitary process $U$ that interacts three systems: our system of interest $A$, the thermometer $B$ and a reference frame (catalyst) system $C$. The total state of these three systems subject to the interaction $U$ is given by $\sigma'_{ABR} := U\left(\rho_A \ot \gamma_B \ot \phi_R \right)U^{\dagger}$, and the heat becomes $Q_R(T, H_B, U)\! :=\! \Tr\left[H_B (\sigma'_B - \gamma_B)\right]$ with $\sigma_B' = \Tr_{AR}\sigma_{ABR}'$. We then define the catalytic cold temperature $T_c^{\circ}(A,\delta)$ as the effective temperature optimized over all possible reference frames under the catalytic constraint, i.e.
\begin{align}
\label{eq:8}
 T_c^{\circ}(A, \delta) :=\! \min_{H_B, U, H_R, \phi_R}   &T 
 \nonumber\\
\textrm{s.t.} \quad & Q_R(T, H_B, U) \leq -\delta, \\
\nonumber
\quad & [U, H_A+H_R+H_B] = 0,\\
& \Tr_{SB}\left[U(\rho_S \ot \gamma_B \ot \phi_R) U^{\dagger}\right] = \phi_R.
\nonumber
\end{align}
The definition for $T_h^{\circ}(A, \mu)$ is found similarly as before, i.e. by replacing $\max$ with $\min$, reversing the side of the inequality and changing $\delta \rightarrow -\delta$. The newly defined catalytic temperatures $T^{\circ}(A,\delta)$ are related with the catalytic temperature for a fixed reference frame $T_{c}(A,\delta|R)$ used in the main text via $T_c^{\circ}(A,\delta) = \min_{R} T_{c}(A,\delta|R)$ and similarly for $T_h^{\circ}(A,\delta)$.

We begin by first showing that the constraints 
\begin{align}
    \label{eq:cat_eff_temp_con1}
    Q_R(T, H_B, U) &\leq -\delta \\
    \label{eq:cat_eff_temp_con2}
    \Tr[U, H_A + H_R + H_B] &= 0, \\
    \label{eq:cat_eff_temp_con3}
    \Tr_{SB}[U(\rho_S \ot \phi_R \ot \gamma_B)U^{\dagger}] &= \phi_R,
\end{align}
imply that $F_T(\rho_A, H_A) \geq F_T(\sigma_A, H_A)$ for any $H_B, H_R, U$ and $\phi_R$. To see this notice that, similarly as before, the unitary invariance of von Neuman entropy implies
\begin{align}
    \Delta S_S + \Delta S_R + \Delta S_B \geq 0.
\end{align}
Our constraints imply $\Delta S_R = 0$ and $\Delta S_B = -\beta \Delta E_S - D(\rho_B'\| \gamma_B) \leq -\beta \Delta E_S$, and therefore we obtain
\begin{align}
    \label{eq:eff_temp_cat_1}
     F_T(\rho_A, H_A) \geq F_T(\sigma_A, H_A),
\end{align}
which is valid for any choice of operators $H_B$, $H_R$, $U$ and $\phi_R$. Moreover, the constraints also imply that $\Delta E_S + \Delta E_B = 0$ and therefore 
\begin{align}
    \label{eq:eff_temp_cat_2}
    \Tr\left[H_A(\sigma_A - \rho_A)\right] \geq \delta.
\end{align}

Let us now assume that Eqs. (\ref{eq:eff_temp_cat_1}) and (\ref{eq:eff_temp_cat_2}) hold. We want to prove that there always exist feasible operators $H_B$, $H_R$, $U$ and $\phi_R$ that achieve the same optimal value as in Eq. (\ref{eq:equiv_problem}--\ref{eq:con_2}). Let us consider
\begin{align}
    \label{eq:def_duan_state}
    \phi_R^n = \frac{1}{n}\sum_{i=1}^{n-1} (\rho^{\ot n} \ot \sigma^{\ot n-i})_{R_A} \ot \dyad{i}_{R_M}, \qquad H_R = H^{\ot {n-1}}_{R_A} \oplus \mathbb{1}_{R_M},
\end{align}
where we denote $R = R_A R_M$, $H_{R_A} = H_A$, $\sigma^{n-i} := \Tr_{1:i}[\sigma^n]$ with $\Tr_{1:i}[\cdot]$ denoting the partial trace over the first $i$ particles and $\sigma^{n}$ an arbitrary $n$-partite density matrix which we will soon specify. The unitary $U$ is chosen to be
\begin{align}
    U := U_{ARB} = W_{AR} \left(\sum_{i = 1}^n U^{(i)}_{AR_AB} \ot \dyad{i}_{R_M}\right), 
\end{align}
where $\{U^{(i)}\}$ is a collection of unitaries chosen such that $U^{(i)} = \mathbb{1}^{\ot n} \ot \mathbb{1}_B$ for $1 \leq i < n$ and $U^{(n)} = V$ which we will specify later. The unitary $W_{AR}$ is a cyclic permutation in the sense that:
\begin{align}
    W_{AR}\left[ \ket{i_1}_{A} \ot \left( \ket{i_2} \ot \ldots \ot \ket{i_n}\right)_{R_A} \ot \ket{i}_{R_M}\right] = \ket{i_n}_{A} \ot \left( \ket{i_1} \ot \ldots \ot \ket{i_{n-1}}\right)_{R_A} \ot \ket{i+1}_{R_M},
\end{align}
with $\ket{n+1}_{R_M} \equiv \ket{1}_{R_M}$. Up to this point we still need to specify $H_B$, $V$ and $\sigma^n$. Notice that $U$ leads to the global state $\sigma_{ARB} = U (\rho_A \ot \phi_R \ot \gamma_B)U^{\dagger}$ with $\sigma_{AR} := \Tr_B\sigma_{ARB}$ of the form
\begin{align}
    \label{eq:sigma_AR}
    \sigma_{AR} &= \frac{1}{n} \Tr_B \! \left[W_{AR}\! \left(\rho_A \ot \sigma^{\ot n-1}_{R_A} \ot \gamma_B \ot \dyad{1}_{R_M} + \ldots + \Tr_B[V(\rho^{\ot n}_{AR_A} \ot \gamma_B)V^{\dagger}] \ot \dyad{n}_{R_M}\right) W_{AR}^{\dagger}\right]. 
\end{align}
Notice that $U$ is energy-preserving as long as $V$ is energy preserving on its support. By the assumption (\ref{eq:con_1}) and Lemma \ref{lem:1} we know that for any $\epsilon > 0$ there is a sufficiently large $n$, a system $B = (H_B, \gamma_B)$ and energy-preserving unitary $V$ acting on the subspace $AR_AB$ such that
\begin{align}
    \label{eq:unitary_from_lemma}
    \norm{\Tr_B\left[V(\rho^{\ot n}_{AR_A} \ot \rho_B)V^{\dagger}\right] - \sigma^{\ot n}_{AR_A}}_1 \leq \epsilon.
\end{align}
As indicated by the notation, we choose the unitary from Eq. (\ref{eq:unitary_from_lemma}) to be the unitary used in the catalytic protocol (see Eq. (\ref{eq:sigma_AR})) and our thermometer system $B$ to be the environment system whose existance is assured by Lemma \ref{lem:1}. This assures that $U$ is energy-conserving, i.e. Eq. (\ref{eq:cat_eff_temp_con2}) holds. Finally, the density matrix $\sigma^n$ is chosen to be
\begin{align}
    \sigma^n := \mathcal{E}_T(\rho^{\ot n}_{AR_A}) = \Tr_B\left[V(\rho^{\ot n}_{AR_A} \ot \rho_B)V^{\dagger}\right].
\end{align}
With these identifications we find
\begin{align}
    \sigma_{AR} &= \frac{1}{n} \left( \sigma^n_{AR_A} \ot \dyad{1}_{R_M} + \sum_{i=2}^{n} (\rho^{\ot i-1} \ot \sigma^{n-i})_{AR_A} \ot \dyad{i}_{R_M}\right).
\end{align}
A simple calculation further shows that
\begin{align}
    \Tr_A[\sigma_{AR}] = \phi_R, \qquad \norm{\sigma_{AR} - \sigma_A \ot \phi_R} \leq \mathcal{O}(\epsilon),
\end{align}
i.e. Eq. (\ref{eq:cat_eff_temp_con3}) is satisfied and moreover, the protocol generates arbitrarily small correlations between the catalyst $R$ and system $A$.  Moreover, denoting with $\Tr_{/i}(\cdot)$ the partial trace over systems $\{1, \ldots, i-1, i+1, \ldots, n\}$ we can verify that $\Tr_R \sigma_{AR} = \sigma_A$ reads
\begin{align}
    \sigma_A = \frac{1}{n} \sum_{i=1}^n \Tr_{/i}\mathcal{E}_T(\rho_A^{\ot n}).
\end{align}
By our construction $\mathcal{E}_T(\rho^{\ot n})$ is close to $\sigma^{\ot n}$, with $\sigma$ being any state satisfying $F_T(\rho, H_A) \geq F_T(\sigma, H_A)$. Using again the fact that $\mathcal{E}(\rho^{\ot n}) \geq \sigma^{\ot n} - \epsilon \mathbb{1}^{\ot n}$ and $\epsilon \propto \mathcal{O}(e^{-an})$, we obtain
\begin{align}
\Tr[H_A(\sigma_A - \rho_A)] &= \frac{1}{n} \sum_{i=1}^n \tr[H_A(\Tr_{/i} \mathcal{E}_T(\rho_A^{\ot n}) - \rho_A)] \\
    &= \frac{1}{n}\Tr[H_A^{\ot n}(\mathcal{E}_T(\rho^{\ot n}_A) - \rho^{\ot n}_A)]\\    
    & \geq \Tr[H_A^{\ot n}\left(\sigma_A^{\ot n} - \rho_A^{\ot n}\right)] - n \epsilon \Tr[H_A] \\
     &\geq n \delta - \mathcal{O}(ne^{-an}).
\end{align}
Since $Q(T, H_B, U) = -\Tr[H_A(\sigma_A - \rho_A)]$, we recover the final constraint (\ref{eq:cat_eff_temp_con1}). Therefore, by taking $n$ sufficiently large, which amounts to considering larger catalytic reference frames in Eq. (\ref{eq:def_duan_state}), the solution of our optimization problem approaches the solution of the simplified problem (\ref{eq:equiv_problem}--\ref{eq:con_2}). In the limit $n \rightarrow \infty$ the two problems therefore achieve the same optimal value of $T$. The best possible temperatures $T^{\circ}_{c/h}$ are therefore achieved only when infinitely large systems are used as reference frames. Moreover, when system $A$ is coherent in the energy basis, the optimal catalyst is necessarily coherent in the energy basis. By limiting ourselves to arbitrarily large but incoherent catalysts we obtain expressions similar to Eqs. (14), with $\rho_A$ replaced by a dephased state $\mathcal{D}(\rho_A) := \sum_{i} \dyad{\epsilon_i} \rho_A \dyad{\epsilon_i}$.
\end{document}